\documentclass[twocolumn,aps,prd,final,reprint,showkeys,showpacs,superscriptaddress,floatfix,nofootinbib,amsmath,amssymb]{revtex4-2}
\usepackage[utf8]{inputenc}

\usepackage{hyperref} 
\usepackage{url}
\usepackage{graphicx}

\usepackage{graphics}
\usepackage{natbib}
\graphicspath{{./Figure}}

\usepackage{dcolumn}
\usepackage{bm}

\usepackage[normalem]{ulem}
\usepackage{color}

\usepackage{soul}

\usepackage{supertabular}
\usepackage{tabularx}

\def\units#1{~\hbox{$\,{\rm #1}$}}
\def\degrees{\hbox{$^\circ$}}

\usepackage{orcidlink}
\usepackage{comment}

\newcommand{\Ef}{\boldsymbol{E}}

\begin{document}

\title{A 3D Monte Carlo calculation of the inverse Compton emission from the Sun and stars in presence of magnetic and electric fields}

\author{M.~N.~Mazziotta\, \orcidlink{0000-0001-9325-4672}}
\email{marionicola.mazziotta@ba.infn.it (M. Nicola Mazziotta)}

\affiliation{Istituto Nazionale di Fisica Nucleare, Sezione di Bari, via Orabona 4, I-70126 Bari, Italy}


\begin{abstract}
The solar steady emission in gamma rays is due to the interactions of Galactic cosmic rays with the solar atmosphere and with the low-energy solar photon field via inverse Compton scattering. The emission is sensitive to the magnetic field nearby the Sun and to the cosmic-ray transport in the magnetic field in the inner solar system. Modeling the inverse Compton emission in the presence of a magnetic field is therefore crucial to better interpret the observations. In a previous work we have presented a comprehensive calculation of the secondary productions due to the collision of cosmic rays with the solar atmosphere in presence of magnetic fields. In this paper, we present a general approach to calculate the (anisotropic) inverse Compton scattering in a 3D Monte Carlo simulation, also in presence of magnetic and electric fields. After a short review of the scattering process of photons with electrons, examples of inverse Compton emission are presented, including the predictions for the Sun.
\end{abstract}

\keywords{Inverse Compton, Sun, Gamma rays, Heliosphere}

\maketitle

\section{Introduction}

Inverse Compton (IC) scattering, i.e. the upscattering of low-energy photons by high-energy electrons, is one of the main mechanisms for the production of X rays and gamma rays in astrophysical environments and in particle beam facilities. Cosmic-ray electrons and positrons (CREs) can scatter with the interstellar radiation field (including starlight and cosmic microwave background radiation), contributing to the Galactic diffuse gamma-ray emission.

The steady emission of the solar disk in gamma rays is due the collisions of Galactic cosmic-ray nuclei and electrons with the solar atmosphere~\cite{Abdo:2011xn,Seckel:1991ffa,Mazziotta:2020uey} via hadronic and bremsstrahlung scattering effects; on the other hand, the collisions of Galactic cosmic-ray electrons (and positrons) with the solar optical photons via IC scattering produce a diffuse component extending up to tens of degrees from the Sun~\cite{Abdo:2011xn,Orlando:2008uk,Orlando:2006zs,Moskalenko:2006ta,Orlando:2020ezh}.

The solar IC emission is currently modeled by integrating along the line of sight from the observer to the Sun the IC spectral intensity, which is evaluated by folding the CRE intensity with the IC scattering cross section on the low-energy photons originated from the black-body radiation produced by the Sun (or more generally by a star)~\cite{Orlando:2008uk,Orlando:2006zs,Moskalenko:2006ta,Orlando:2020ezh,Lai:2022qif}.

In our previous work~\cite{Mazziotta:2020uey} we have calculated the disk emission by simulating the interactions of CRs with the Solar atmosphere using the {\tt FLUKA} simulation package, and we found that this emission depends on the magnetic field nearby the Sun and on the cosmic-ray transport in the magnetic field in the inner solar system. However, the predicted flux is lower than observations above tens of GeV, even in the case of strong magnetic field configurations nearby the Sun. In particular, the disk emission data suggest a harder spectrum ($\sim E_\gamma^{-2.2}$) than the cosmic-ray spectrum ($\sim E_{p}^{-2.7}$). On the other hand, the IC gamma-ray intensity is expected to follow a power-law $\sim E_\gamma^{-(\alpha+1)/2}$ with $\alpha$ the spectral index of CRE intensity~\cite{ginzburg1964gamma}. Since $\alpha\approx 3$ for the local CRE~\cite{Fermi-LAT:2009yfs,Fermi-LAT:2009ppq,Aguilar:2014fea,Fermi-LAT:2017bpc,Ambrosi:2017wek} the IC spectrum is expected to have a spectrum $\sim E_\gamma^{-2}$.

The magnetic field is also expected to affect the IC gamma-ray emission close to the Sun, since the electrons should move along curved trajectories, whose lengths determine the interaction probability with the intense optical photon field. The IC emission is therefore expected to be brighter towards the Sun direction, and might be not well separated from the disk emission. In this context, to get a complete picture of the solar gamma-ray emission, the IC scattering should also be calculated in presence of strong and irregular magnetic field. 

The IC emission spectrum formed through multiple scattering by electrons has been calculated with Monte Carlo simulations in several previous papers (see for example   \cite{1976SvAL....2...55P,1977SvA....21..708P,1979A&A....75..214P}). In this paper, we present a general approach to calculate the (anisotropic) IC scattering for a 3D Monte Carlo problem, also in presence of magnetic and electric fields. 

In Sec.~\ref{sec:icformalism} we first describe the formalism for the differential cross section, based mainly on standard textbooks (see for instance~\cite{berestetskii1982quantum}). In Sec.~\ref{sec:MCICheadon} we present a Monte Carlo approach for head-on collision calculation using a fixed electron beam energy with a given set of photon energies. In Sec.~\ref{sec:ICiso} we apply the Monte Carlo approach to the IC scattering with the isotropic black-body thermal photons produced by Galactic CREs. In Sec.~\ref{sec:ICstar} we describe our calculation of the effective IC cross section due to the scattering of electrons near a star, and we show the numerical results using the Sun as test case. Then, in Sec.~\ref{sec:MCICstar} we evaluate the IC spectral emission still using the Sun as test case, either without or with magnetic and electric fields. Finally, we conclude and discuss the outlook of this work in Sec.~\ref{sec:conc}.

\section{Scattering of a photon by an electron}
\label{sec:icformalism}

The law of four-momentum conservation in the scattering of a photon by a free electron (the Compton effect) is given by:

\begin{equation}
    p + k = p' + k'
\end{equation}
where $p$ ($p'$) and $k$ ($k'$) are the 4-momenta of the electron and of the photon before (after) the collision. The kinematic invariants are:

\begin{eqnarray}
s &= (p+k)^2  =  (p'+k')^2  =  m^2+2pk=m^2+2p'k' \nonumber \\
t &= (p-p')^2 = (k'-k)^2 =  2(m^2-pp')=-2kk' \\
u &= (p-k')^2 = (p'-k)^2 =  m^2-2pk' = m^2-2p'k \nonumber
\end{eqnarray}
where $m$ is the electron mass.

The unpolarized cross-section is given by~\cite{berestetskii1982quantum}:

\begin{eqnarray}
d\sigma_C & = 8 \pi r_e^2 \frac{m^2~dt}{(s-m^2)^2} \Biggr\{
\left( \frac{m^2}{s-m^2} + \frac{m^2}{u-m^2} \right) ^2 + \nonumber \\
& \left( \frac{m^2}{s-m^2} + \frac{m^2}{u-m^2} \right) - \frac{1}{4} \left( \frac{s-m^2}{u-m^2} + \frac{u-m^2}{s-m^2} \right) \Biggr\}
\end{eqnarray}
where $r_e$ is the classical radius of the electron.

Taking into account that $s+u+t=2m^2$, from four-momentum conservation $ds=0$ and therefore $dt=-du$. The allowed values of $u$ are in the range $m^4/s \leq u \leq 2 m^2 -s$.

Expressing $s$ and $u$ in terms of the quantities:

\begin{align}
    x = & \frac{s-m^2}{m^2} \nonumber \\
    y = & \frac{m^2-u}{m^2} 
\end{align}
and taking into account that:
\begin{equation}
    dt = -du = m^2 dy
.\end{equation}
we obtain~\cite{berestetskii1982quantum}:

\begin{eqnarray}
    \frac{d\sigma_C}{dy} && = \frac{8 \pi r_e^2}{x^2} \left[ \left( \frac{1}{x} - \frac{1}{y}\right)^2 + \left( \frac{1}{x} - \frac{1}{y}\right) + \frac{1}{4} \left( \frac{x}{y} + \frac{y}{x}\right) \right] \nonumber \\ && = \frac{8 \pi r_e^2}{x^2} f(x,y)
    \label{eq:sigC}
\end{eqnarray}
with $x/(1+x) \leq y \leq x$. In the lab frame $x$ and $y$ are given by:

\begin{align}
    x = & \frac{2 ~ E_e ~ \epsilon ~ (1-\beta \cos \theta)}{m^2} \nonumber \\
    y = & \frac{2 ~ E_e ~ E_\gamma ~ (1-\beta \cos \theta_1)}{m^2} 
    \label{eq:xylab}
\end{align}
where $\theta$ is the angle between the electron with initial energy $E_{e}$ and the incoming photon with energy $\epsilon$, while $\theta_1$ is the angle between the initial direction of the electron and the scattered photon with energy $E_\gamma$;
$\beta$ and $\gamma$ are given by:

\begin{equation*}
    \beta = \frac{|\vec{p}|}{E_e} , 
    \gamma = \frac{1}{\sqrt{1-\beta^2}}
\end{equation*}
where $\vec{p}$ is the initial momentum of the electron.

Introducing the energies of the photon in the electron rest frame (ERS) before and after the scattering, $\omega$ and $\omega'$, respectively, defined as:

\begin{align}
    \omega = & \frac{E_e ~ \epsilon ~ (1-\beta \cos \theta)}{m} \nonumber \\
    \omega' = & \frac{E_e ~ E_\gamma ~ (1-\beta \cos \theta_1)}{m} 
    \label{eq:omlab}
\end{align}
we obtain $x=2 \omega/m$ and $y=2 \omega'/m$. 

Since $s+u+t=2m^2$ and $t=-2kk'=-2 \epsilon E_\gamma (1-\cos \eta)$, where $\eta$ is the angle between the scattered photon and the incoming photon, we get:

\begin{equation}
m^2(x+1)+m^2(1-y)+t=2m^2 \Longrightarrow t=m^2(y-x).    
\label{eq:ttlab}
\end{equation}
Combining Eqs.~\ref{eq:xylab} and~\ref{eq:ttlab} for $t$, we get $2 \epsilon E_\gamma (1-\cos \eta)=2 ~ E_e ~ \epsilon ~ (1-\beta \cos \theta) - 2 ~ E_e ~ E_\gamma ~ (1-\beta \cos \theta_1)$. The energy of the scattered photon $E_\gamma$ is therefore given by:

\begin{eqnarray}
    E_\gamma &= \frac{E_e ~ \epsilon ~ (1-\beta \cos \theta)}{E_e ~ (1-\beta \cos \theta_1) + \epsilon (1-\cos \eta)} = \Phi(E_e, \epsilon, \theta, \theta_1, \eta )
    \label{eq:Phi}
\end{eqnarray}
and is a function of the initial electron and photon energies $E_e$ and $\epsilon$, and of the angles $\theta$, $\theta_1$ and $\eta$.

In a laboratory frame in which the incident electron is moving along the z-axis direction, the incident photon is propagated along the direction given by the polar angle $\theta$ and the azimuthal angle $\varphi$. After the collision, the photon is scattered into the direction given by the polar angle $\theta_1$ and the azimuthal angle $\phi_1$. Therefore the angle between the momentum of the incident photon and that of the scattered photon $\eta$ is given by:

\begin{equation}
    \cos \eta = \cos \theta \cos \theta_1 + \sin \theta \sin \theta_1 \cos (\phi-\phi_1).
\end{equation}

Replacing $dy$ in Eq.~\ref{eq:sigC}, the angular differential cross section is given by

\begin{equation}
    \frac{d\sigma_C}{d\Omega_1} = \frac{8 r_e^2 E_\gamma^2}{m^2 x^2} f(x,y) 
    \label{eq:xsang}
\end{equation}
where we used~\cite{akhiezer1965quantum}

\begin{equation}
    dy = \frac{E_\gamma^2}{\pi m^2 x^2} d\Omega_1.
\end{equation}

In the ERS, the electron initial four-momentum is given by $p^*=(m, \vec{0})$ and the photon initial four-momentum is given by $k^*=(\omega^*,\vec{k}^*)$ with $\omega^*=k^*$. In this frame $s-m^2=2 m \omega^*$, $u-m^2=-2m\omega'^{*}$ and $t=-2kk'=-2 \omega^* \omega'^* (1-\cos \eta$), where $\omega'^*$ is the energy of the photon after the collision. In this reference system we also have: $x=2 \omega^*/m$ and $y=2 \omega'^*/m$. The energy of the scattered photon is given by:

\begin{align}
    \omega'^* = \frac{m \omega^*}{m + \omega^*(1-\cos \eta)} \nonumber \\
    \frac{1}{\omega'^*} - \frac{1}{\omega^*} = \frac{1}{m} (1-\cos \eta).
    \label{eq:omep}
\end{align}

The differential Compton scattering cross section in which the electron is at rest (we are dropping the $^*$ in the following equations), aka Klein-Nishina cross section~\cite{1929ZPhy...52..853K}, is given by ($dt = -du = 2 m d \omega' = 2 (\omega')^2 d\cos \eta = (\omega')^2 ~d \Omega/\pi$ since $d \omega' / (\omega')^2=d \cos \eta/m$ from Eq.~\ref{eq:omep}~\footnote{$t=-2 \omega \omega' (1-\cos \eta$), so $dt=- 2 \omega d \omega' (1-\cos \eta) + 2 \omega \omega' d \cos \eta $}):

\begin{equation}
    \frac{d \sigma_C}{d \Omega} = \frac{r_e^2}{2} \left( \frac{\omega'}{\omega} \right)^2 \left( \frac{\omega}{\omega'} + \frac{\omega'}{\omega} - \sin^2 \eta \right)
\end{equation}
where $d \Omega = 2 \pi \sin \eta d \eta$, with the relation between the the photon energy change and the scattering angle:

\begin{equation}
      \frac{1}{y} - \frac{1}{x} = \frac{1}{2} (1-\cos \eta).
  \label{eq:Cangle}
\end{equation}

Since the angle $\eta$ is unambiguously related to $\omega$, the cross section can be expressed in terms of the energy of the scattered photons $\omega'$~\cite{berestetskii1982quantum}:

\begin{eqnarray}
    \frac{d \sigma_C}{d \omega'} && = \pi r_e^2 \frac{m}{\omega^2} \\ \nonumber 
    && \left[ \frac{\omega}{\omega'} + \frac{\omega'}{\omega} + \left( \frac{m}{\omega'}-\frac{m}{\omega} \right)^2 - 2 m \left( \frac{1}{\omega'} - \frac{1}{\omega} \right) \right]
\end{eqnarray}
with $\omega'$ in the range $\omega/(1+2\omega/m) \leq \omega' \leq \omega$.

The total cross-section for a fixed value of $x$ is given by:
\begin{equation}
    \sigma_C(x) = \int_{x/(1+x)}^x \frac{d\sigma_C}{dy} dy = \sigma_T f(x)
\end{equation}
where $\sigma_T$ is the Thomson cross section $\sigma_T=\frac{8 \pi }{3} r_e^2$ and $f(x)$ is given by: 

\begin{eqnarray}
    \label{eq:fx}
    &f(x) & =
    \frac{3}{4x} \\ \nonumber
    && \left[ \left( 1-\frac{4}{x}-\frac{8}{x^2} \right)  \log(1+x)+\frac{1}{2}+\frac{8}{x}-\frac{1}{2(1+x)^2}
    \right]
\end{eqnarray}

\begin{figure}[!t]
    \centering
    \includegraphics[width=0.48\textwidth, height=0.23\textheight]{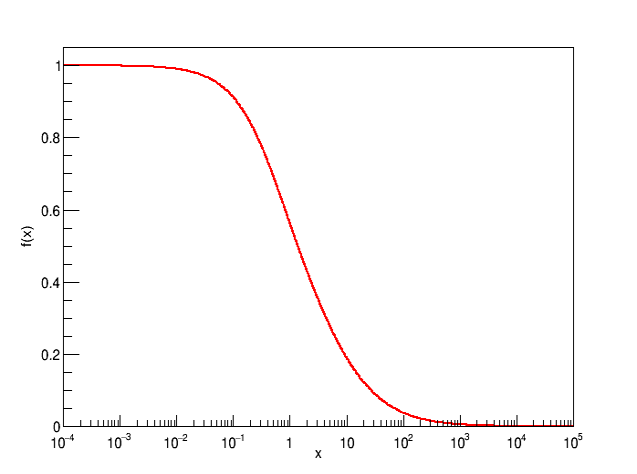}
    \caption{$f(x)$ as a function of x (see Eq.~\ref{eq:fx}).}
    \label{fig:fx}
\end{figure}

The leading term of Eq.~\ref{eq:fx} if $x \ll 1$ (i.e. non relativistic case) is $f(x) \sim 1 -x$, and the first term is the classical Thomson cross section of $\sigma_C(x)=\sigma_T$. In the ultra-relavistic case ($x \gg 1$) the expansion of Eq. \ref{eq:fx} gives $f(x)=\frac{3}{4x} \left( \log x + \frac{1}{2} \right)$~\cite{berestetskii1982quantum}. Fig.\ref{fig:fx} shows the function $f(x)$. 

\section{A Monte Carlo approach in head-on collisions}
\label{sec:MCICheadon}

The scattering of electrons on low-energy  photons (e.g. thermal photons) produces high-energy gamma rays. This process can be simulated by means of a Monte Carlo method.

The simulation starts from the initial value of the electron and photon energy and momenta in the lab system; the photon is then boosted in the ERS, where the collision is simulated according to the equations in the previous section; finally, the scattered photon is boosted back in the lab frame.

The Compton scattering simulation is performed in the ERS assuming that the incoming photon direction is along the z-axis. The scattered photon energy is sampled using a combined composition and rejection Monte Carlo method (see for instance~\cite{1960NucPh..20...15B}), then the Compton angle (corresponding to the polar angle) is calculated. The azimuth angle is generated with a uniform probability in the range $[0, 2\pi[$. 
Finally, the scattered photon is boosted back into the lab system. 

\begin{figure}[!t]
    \centering
    \includegraphics[width=0.49\textwidth, height=0.22\textheight]{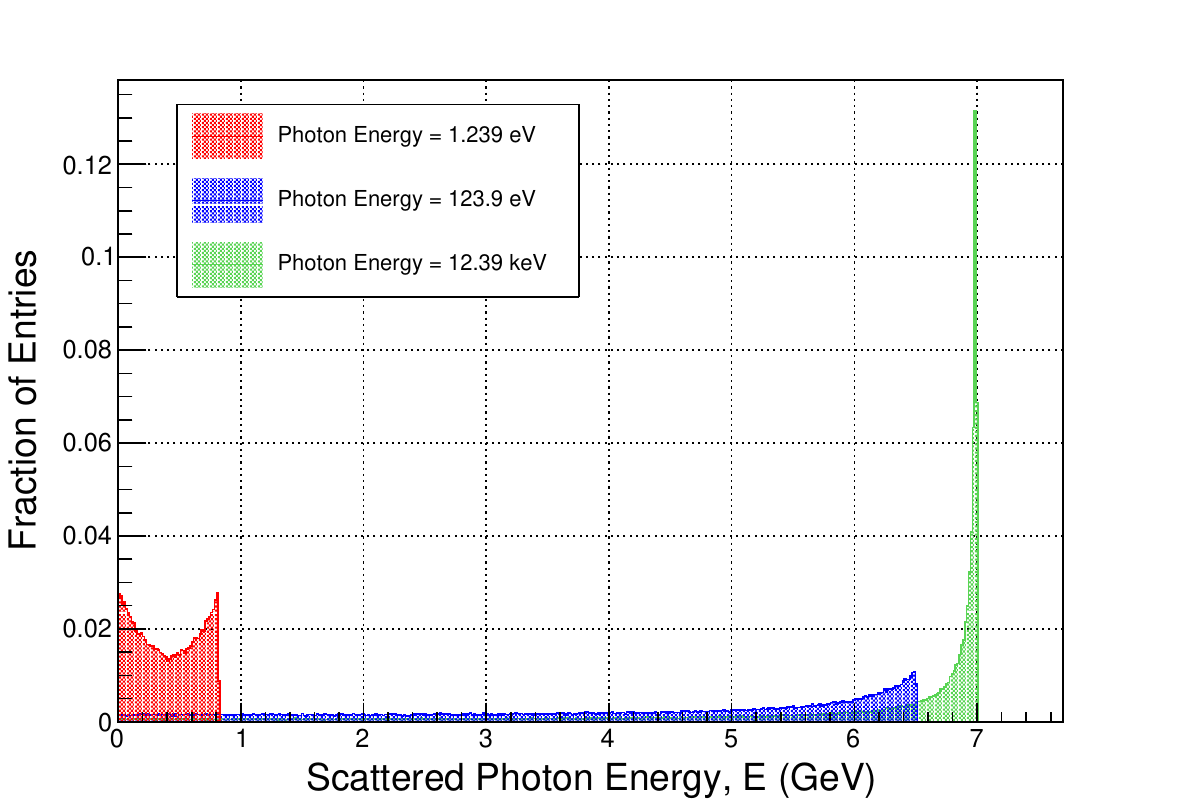}
    \includegraphics[width=0.49\textwidth, height=0.22\textheight]{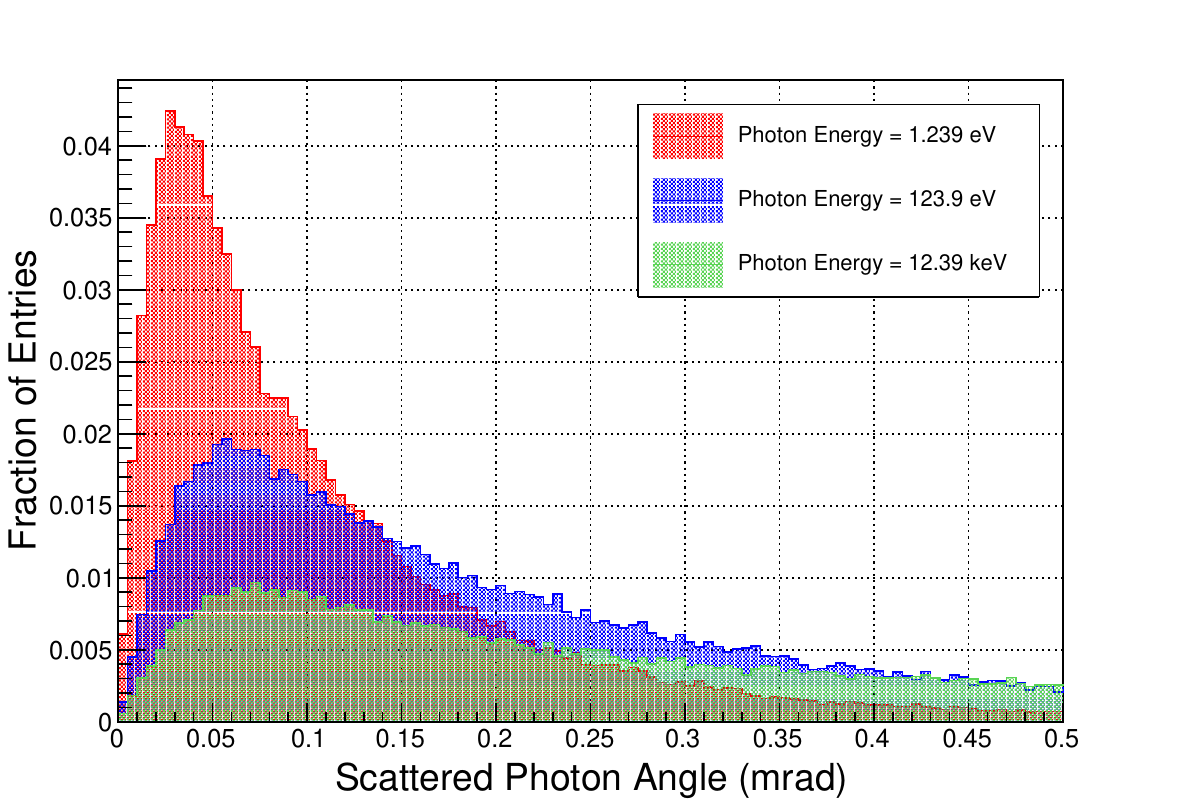}
    \caption{Energy (top panel) and angular (bottom panel) distributions of the emitted photons for head-on collision with 7 GeV electrons. Red: incoming photon energy of 1.239 eV; Blue: incoming photon energy of 123.9 eV; dark green: incoming photon energy of 12.39 keV (see ref.~\cite{Curatolo:2017fme}).}
    \label{fig:IClaser}
\end{figure}

Figure~\ref{fig:IClaser} shows the distribution of energies and angles of the scattered photons in the lab frame in head-on collisions (i.e. $\theta=\pi$) between electrons of 7 GeV kinetic energy and low-energy photons of 1.239 eV, 123.9 eV and 12.39 keV, respectively ~\cite{Curatolo:2017fme}. Our results agree with the ones reported in ref.~\cite{Curatolo:2017fme}.

\section{IC scattering with isotropic thermal photons}
\label{sec:ICiso}

The intensity of the gamma rays, $I(E_{\gamma})$, produced in the interactions of CREs with a radiation field is~\cite{ginzburg1964gamma}:


\begin{equation}
I(E_{\gamma})=\int dE_e ~Y(E_\gamma,E_e)~I_e(E_e) 
\end{equation}
where $I_e(E_e)$ is the intensity of the CREs with energy $E_e$ and $Y(E_\gamma,E_e)$ is the the differential yield of the IC photons with energy $E_\gamma$ per electron given by: 
\begin{equation}
    Y(E_\gamma,E_e)=L \int d\epsilon ~ n_{ph}(\epsilon) ~ \sigma(E_\gamma, \epsilon, E_e)
\end{equation}
where $n_{ph}(\epsilon)$ is the density of photons per unit of volume and energy $\epsilon$, $L$ is the total length along the line of sight and $\sigma(E_\gamma, \epsilon, E_e)$ is the effective cross section for an electron with energy $E_e$ to generate a photon with energy $E_\gamma$ when scattering with a thermal photon with energy $\epsilon$. 

The effective cross section is given by:

\begin{equation}
    \sigma(E_\gamma, \epsilon, E_e) = \int (1 - \beta \cos \theta) ~ d \sigma_C
    \label{eq:xsc}
\end{equation}
where $\theta$ is the angle between the electron and the thermal photon and $d \sigma_C$ is the cross section of the Compton scattering (see above). 
The term $(1 - \beta \cos \theta)$ is the correction factor for non-parallel photon target and electron source.
Note that the quantity $\int d\epsilon ~ n_{ph}(\epsilon) ~ \sigma(E_\gamma, \epsilon, E_e)$ is the average number of secondary photons (i.e. IC scattered photons) per unit length, i.e. this is the IC emissivity per unit length.

For the black-body radiation field, the density of photons per unit volume and unit energy is given by:

\begin{equation}
    n_{ph}(\epsilon) = \frac{1}{\pi^2 (\hbar c)^3} \frac{\epsilon^2}{e^{\epsilon/kT}-1} 
    \label{eq:bbr}
\end{equation}
where $\hbar c \simeq 197.327 \times 10^{-16}\units{GeV~cm}$, $k \simeq 8.617 \times 10^{-14}\units{GeV/K}$ and $\epsilon$ in units of\units{GeV}. The average photon energy $\bar{\epsilon}$ is about 2.7 $kT$ while the total density of photons for unit volume is $\Gamma(3) \zeta(3) (k T/\hbar c)^3/\pi^2  \simeq 2.4 (k T/\hbar c)^3/\pi^2$  where $\Gamma$ is the gamma function and $\zeta$ is the Riemann zeta function.

The cross section in Eq.~\ref{eq:xsc} is defined by the expression:

\begin{eqnarray}
    && \sigma(E_\gamma, \epsilon, E_e) = \\ \nonumber
    && \int (1 - \beta \cos \theta) ~ \frac{d \sigma_C}{d \Omega_1} ~ \delta(E_\gamma - \Phi) ~ P(\Omega) ~ d \Omega ~ d \Omega_1
\end{eqnarray}
where $\Phi$ is given by eq.~\ref{eq:Phi}, $\frac{d \sigma_C}{d \Omega_1}$ is given by eq.~\ref{eq:xsang} and $P(\Omega)=P(\theta, \phi)$ is the angular distribution of the incident photons (we are assuming that the electrons are moving along the z-axis). For an isotropic photon distribution $P(\Omega)=1/4 \pi$.

For a given electron energy $E_e$ and for given incoming photon angle $\theta$ with respect to the electron, the cross section $\sigma(E_e, z)$  (in units of $cm^{-1}$) is obtained by integrating on the all the final states of the scattered photon and is given by:

\begin{equation}
   \sigma(E_e, z) = \int d\epsilon ~ n_{ph}(\epsilon) ~ (1 - \beta z) ~ \sigma_C(x)
   \label{eq:eq24}
\end{equation}
where $z=\cos \theta$ and $x$ is given from Eq.~\ref{eq:xylab}. In the Thomson limit and in the case of head-on collision Eq.~\ref{eq:eq24} becomes $\sigma(E_e, z=-1) = \sigma_T ~ (1 + \beta) \int d\epsilon ~ n_{ph}(\epsilon)$.

The total cross section $\sigma(E_e)$  (also in units of $cm^{-1}$) is given by:

\begin{equation}
   \sigma(E_e) = \int_{-1}^{+1} dz ~ \sigma(E_e, z) ~ P(z)
\end{equation}

\begin{figure*}[!t!h]
    \centering
    \includegraphics[width=0.49\textwidth, height=0.23\textheight]{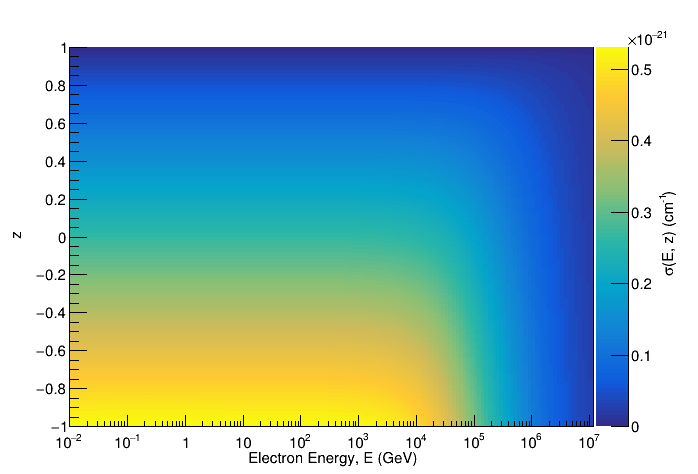}
    \includegraphics[width=0.49\textwidth, height=0.23\textheight]{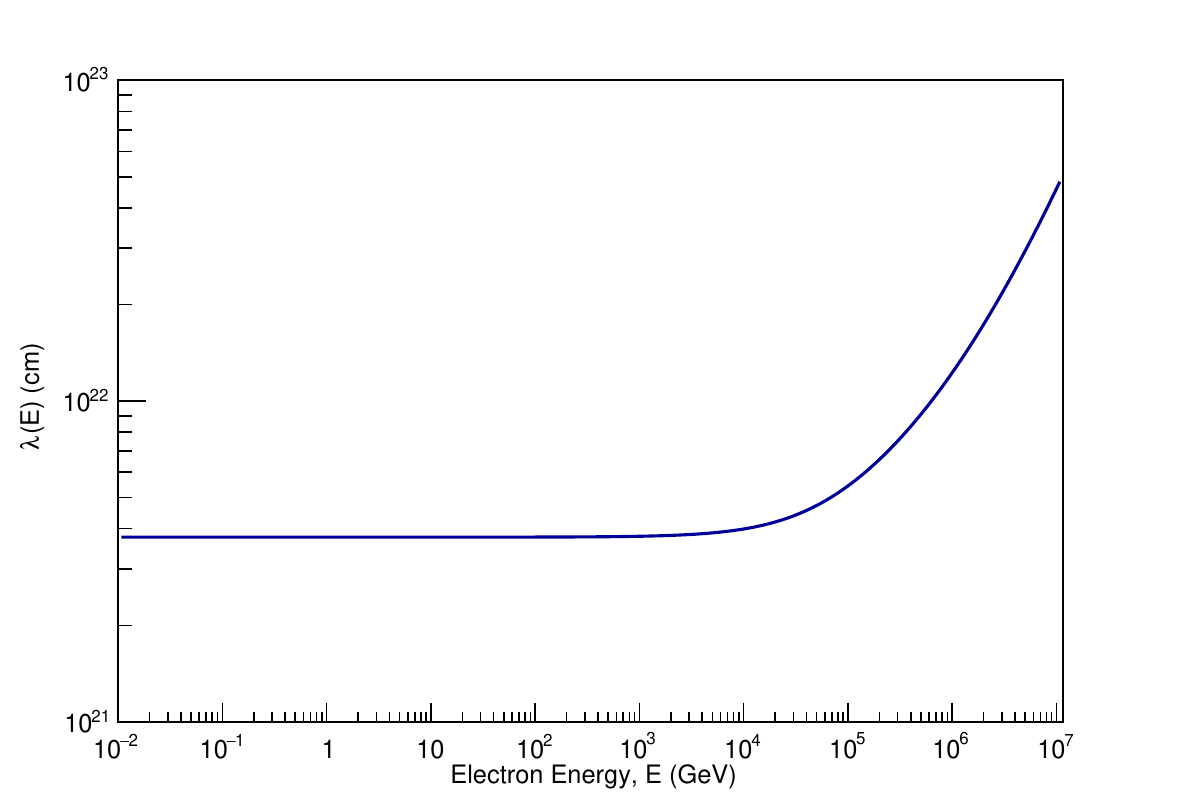}
    \includegraphics[width=0.49\textwidth, height=0.24\textheight]{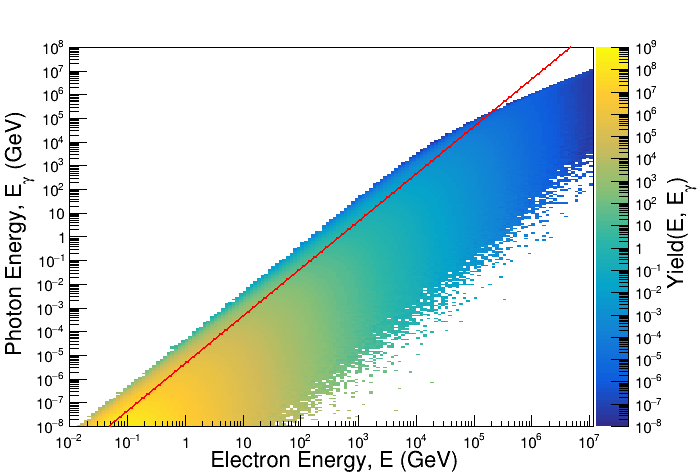}
    \includegraphics[width=0.49\textwidth, height=0.24\textheight]{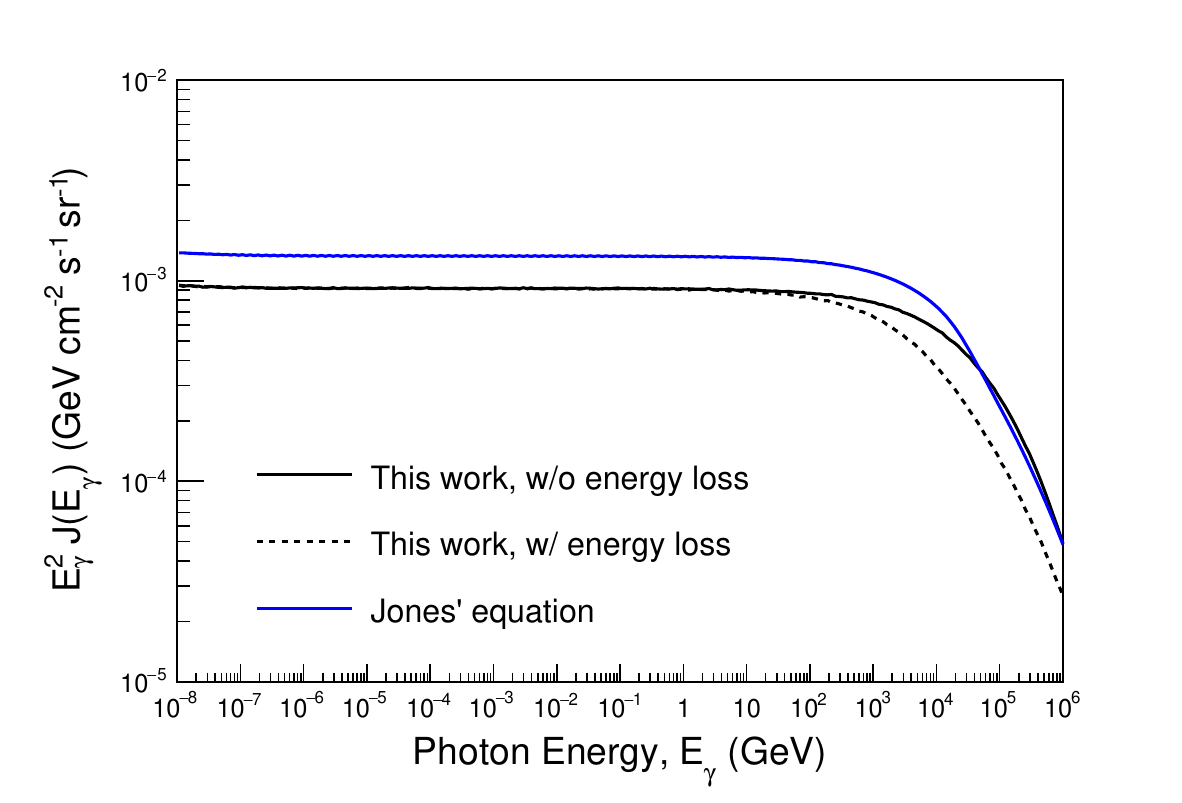}
   \caption{Top left panel: Effective cross section with isotropic photons at 2.7 K as a function of the electron energy and $z=\cos \theta$. Top right panel: Interaction length with isotropic photons at 2.7 K as a function of electron energy.
   Bottom left panel: Inverse Compton photon yield produced in the interactions with isotropic thermal photons at 2.7 K (without the electron energy loss in the scattering). The red line shows the function $E_{\gamma} = 4 \bar{\epsilon} (E_e/m_e)^2/3$, where $\bar{\epsilon} \simeq 2.7 k T$. Bottom right panel: Inverse Compton photon intensity produced by a source of electrons with a power-law energy spectrum $I_e(E)=10^2 (E_e/1 GeV)^{-3}$, scattered by isotropic thermal photons at 2.7 K. Black line: this work, with a Monte Carlo calculation of the IC yield, without (solid line) and with (dashed line) the electron energy loss in the scattering; blue line: IC yield calculated with the Jones formula~\cite{Jones:1968zza}. The total electron path length in the isotropic photon field is assumed to be 10 kpc.
   }
    \label{fig:iso}
\end{figure*}

Thus, for a given electron energy $E_e$, the total interaction length for the IC scattering $\lambda(E_e)$ in a low energy photon field is given by:

\begin{equation}
   \lambda(E_e) = \frac{1}{\sigma(E_e)}
\end{equation}

For an isotropic photon distribution $P(z)=\frac{1}{2}$ and:

\begin{equation}
\sigma(E_e) = \frac{1}{2} \int_{-1}^{+1} dz ~ \int d\epsilon ~ n_{ph}(\epsilon) ~ (1 - \beta z) ~ \sigma_C(x) 
\end{equation}

The differential yield of IC photons can be calculated with a Monte Carlo method. For an electron with energy $E_e$ at the initial position $\vec{r}_0$, the probability density function (PDF) for the path length $s$ from the current position to the site of the next collision is $P(s)=\lambda^{-1} ~ e^{-s/\lambda}$. Random values of $s=s_r$ are generated by using the sampling formula $s_r= - \lambda \log(1-u)$, where $u$ is a random number distributed with a uniform PDF in $[0,1[$. The following interaction occurs at the position $\vec{r}_{i+1} = \vec{r}_{i} + s \hat{d}_{i}$, where $\hat{d}_{i}$ is the direction unit vector (if the initial direction is along the z-axis, $\hat{d}_{0}=(0, 0, 1)$). This procedure continues until the total step length reaches its maximum value $L$ (or the electron leaves the spatial region to be simulated). 

To increase the tracking resolution in the simulation, it is necessary to limit the length $s$ below a given maximum value $s_{max}$ ~\cite{penelope}. To accomplish this, we still sample the free path length $s_r$ to the next interaction from the exponential PDF, but when $s_r > s_{max}$ we only let the particle advance for a distance $s_{max}$ along the direction of motion (i.e., the particle keeps its energy and direction unchanged). On the other hand, when $s_r < s_{max}$, a real interaction is simulated. 
For each real interaction, we randomly extract the energy of the thermal photon and its (isotropic) angle w.r.t. the electron direction in the lab system. From the knowledge of $E_e$, $\epsilon$ and $\theta$, we calculate the secondary IC photon energy and its direction in the  ERS system, and then we boost them back in the lab system (see Sec.\ref{sec:MCICheadon}). The simulation is repeated multiple times for each electron energy. 

The yield of secondary IC photons produced from electrons, $Y(E_{\gamma} | E_e)$, is calculated by counting the secondary IC photons produced with the above Monte Carlo approach. The yield is defined as:

\begin{equation}
Y(E_{\gamma} |~ E_e) = \frac{N(E_{\gamma} |~ E_e)}
{N(E_e) \Delta E_{\gamma}}
\label{eq:yield}
\end{equation} 
where $N(E_e)$ is the number of electron primaries with kinetic energy $E_e$ and $N(E_{\gamma}|~E_e)$ is the number of IC secondary photons with energies between $E_{\gamma}$ and $E_{\gamma} + \Delta E_{\gamma}$.

Figure~\ref{fig:iso} shows the effective cross section (top left panel) and the interaction length (top right panel) for IC scatterings on an isotropic distribution of photons of 2.7 K. Fig.~\ref{fig:iso} also shows the yield of the scattered photons as a function of the electron energy (bottom left panel) and the photon spectrum folded with an electron intensity spectrum following a power-law with spectral index -3 (bottom right panel). These results have been obtained either neglecting the energy losses of electrons (black solid line in the bottom right plot, i.e. each electron can scatter only once), or including the electron energy losses (dashed black line). The blue line shows the gamma-ray spectrum evaluated by using the IC yields calculated with the Jones formula \cite{Jones:1968zza}. The slight difference between the results from the Monte Carlo calculation and those from the Jones formula could be ascribed to the finite width of the energy bins.~\footnote{It is worth to point out that in the simulation we generated the electron energy with a uniform PDF in each bin and the IC photon yield has been calculated in finite photon energy bins, while the Jones formula has been implemented considering the geometric average energy in each bin.}. We have simulated a photon field geometry consisting of an infinite slab with a thickness $L=10$ kpc along the z-axis. The maximum step $s_{max}$ has been set either either to $10\%$ or $1\%$ of the free path length (which depends on the electron energy), yielding the same result. At high energies, the energy loss of electrons should be taken into account, since the scattering becomes much harder with a significant fraction of energy lost by electrons in individual interactions. 

\section{Inverse Compton emission from a star}
\label{sec:ICstar}

In this case the thermal photons are anisotropic, since they are emitted from an extended source. The thermal photon intensity depends on the distance with respect to the star position. 

\begin{figure}[!t!h]
\centering
    \includegraphics[width=\columnwidth]{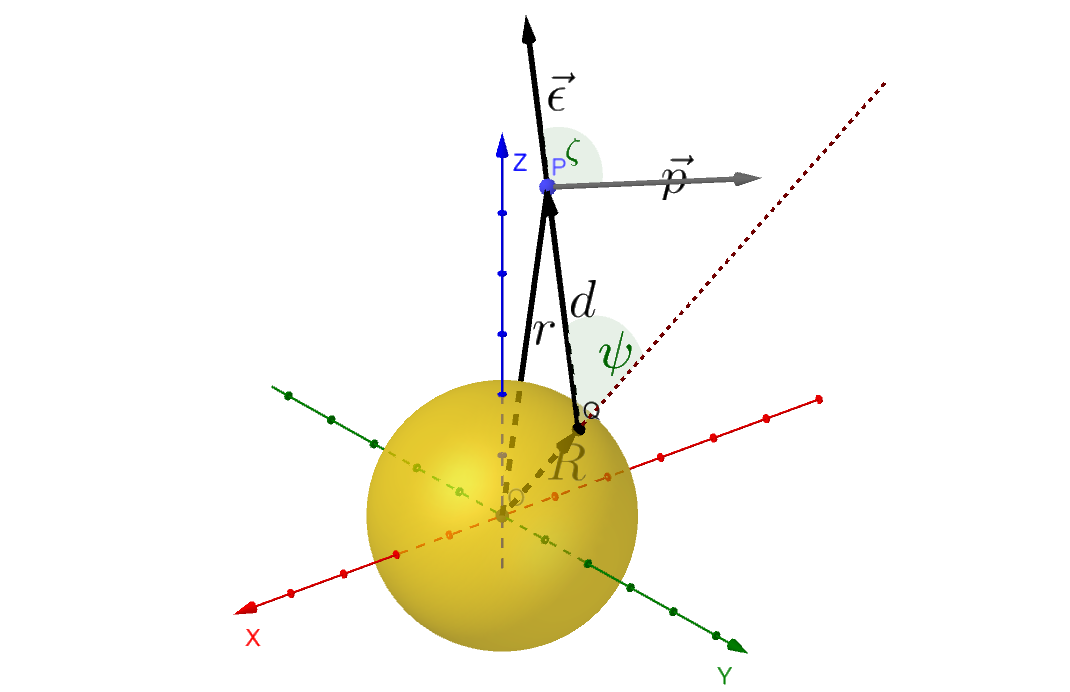}
    \includegraphics[width=\columnwidth]{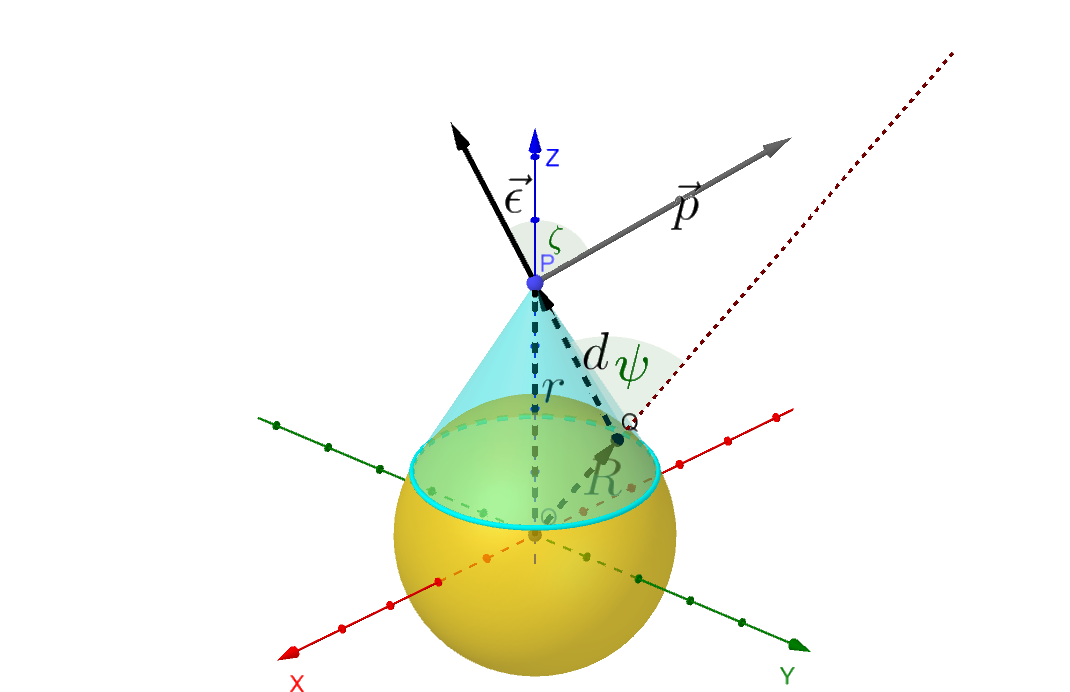}
    \caption{Definition of variables describing the geometry of the inverse Compton emission, in the lab reference system (master reference system) centered with the star position.
    Top panel: diagram for the IC scattering between a photon and an electron in the master reference coordinate system.
    Bottom panel: same as the top panel, with the assumption that we can choose the electron position on the Z-axis. 
    The plots have been made with the  {\tt GeoGebra} tool~\cite{geogebra}.}
    \label{fig:SunIC3D}
\end{figure}

Fig.~\ref{fig:SunIC3D} shows the variables describing the geometry in the lab reference  system (master reference system) centered with the star (e.g. Sun) position. The notation is the following:

\begin{itemize}
    \item $O$ is the center of the star;
    \item $\varphi$ and $\theta$ are the azimuthal and polar angles in the star centered reference system respectively;
    \item $R$ is the star radius;
    \item $Q$ is an emission point on the star surface, i.e. $\overrightarrow{OQ}=(R\cos\varphi \sin \theta, R \sin\varphi \sin\theta, R \cos\theta)$
    \item $dA$ is the area of the surface element of the star at Q emitting photons, $dA=R^2 \sin\theta ~d\varphi~d\theta$;
    \item $P$ is the position of the electron in the star centered reference system;
    \item $r$ is the distance of the electron from the star center, i.e. $r=| \overrightarrow{OP} |$;
    \item $\bar{\theta}$ is the aperture angle of the cone tangent to the star with the vertex at $P$, i.e. $\cos \bar{\theta}= R/r$;
    \item $\vec{p}$ is the electron momentum, $\vec{p}=(p_x, p_y, p_z)=p~ (c_x, c_y, c_z)= p ~ \hat{p}$;
    \item $\vec{\epsilon}$ is the solar photon momentum, $\vec{\epsilon}= (k_x, k_y, k_z)=\epsilon ~\hat{\epsilon}$;
    \item $\zeta$ is the angle between the electron and incoming photon, $\cos \zeta = \frac{\vec{p} \cdot \vec{\epsilon}}{|\vec{p}| |\vec{\epsilon}|}$;
    \item $d$ is the distance between the photon emission point and the electron position, i.e. \\ $d=| \overrightarrow{QP} | = \sqrt{ R^2+r^2 - 2 ~R~ r~ \cos \theta}$ = $r \sqrt{1 + \rho^2 - 2 \rho \cos \theta}$ \\ with $\rho=R/r$;
    \item $\psi$ is the angle of the emitted photons w.r.t. the local zenith at the point $Q$, i.e. \\ $\cos\psi = \frac{\overrightarrow{QP}}{|\overrightarrow{QP}|} \cdot \frac{\overrightarrow{OQ}}{|\overrightarrow{OQ}|}=\frac{r\cos \theta-R}{d}=\frac{\cos \theta -\rho}{\sqrt{1 + \rho^2 - 2 \rho \cos \theta}}$.
\end{itemize}

Following the above notations, the effective cross section (in unit of $cm^{-1}$) for an electron with a given kinetic energy $E_e$ and direction $\hat{p}$ travelling at the distance $r$ from the center of the star, is given by:

\begin{equation}
   \sigma(E_e, r, \hat{p}) = \int dn ~ (1-\beta z)~\sigma_{C}(x)
   \label{eq:eq30}
\end{equation}
where $z=\cos{\zeta}$, $x=\frac{2\epsilon E_e}{m_e^2}(1-\beta z)$ and $dn = n_{ph}(\epsilon)~ d\epsilon~dA~\frac{\cos{\psi}}{4 \pi d^2} = n_{ph}(\epsilon)~ d\epsilon~R^2~\sin{\theta}~ d\theta~d\varphi~\frac{\cos{\psi}}{4 \pi d^2}$. The $\cos \psi$ factor takes into account the density radiation per unit of area of the emitting surface (Lambert's cosine law). The effective cross section is:

\begin{eqnarray}
 && \sigma(E_e, r, \hat{p})  = \frac{R^2}{4 \pi} \int d\varphi \times \nonumber \\ 
 && \int_0^{\bar{\theta}} d\theta \frac{\sin\theta ~ \cos \psi}{d^2} \int d\epsilon~ n_{ph}(\epsilon)~ (1-\beta z)~ \sigma_C(x) = \nonumber \\
 && \frac{R^2}{4 \pi} \int d\varphi \times \nonumber \\ 
 &&\int_{R/r}^{1} d\mu \frac{r \mu - R}{d^3} \int d\epsilon~ n_{ph}(\epsilon)~ (1-\beta z)~ \sigma_C(x) = \nonumber \\
  && \underbrace{\frac{\rho^2}{4 \pi} \int d\varphi \int_{\rho}^{1} d\mu \frac{\mu - \rho}{(1 + \rho^2 - 2 \rho \mu)^{3/2}}}_\text{Spatial term} \times \nonumber \\
  &&\underbrace{\int d\epsilon~ n_{ph}(\epsilon)~ (1-\beta z)~ \sigma_C(x)}_\text{Emission-Kinematic term}
\end{eqnarray}
where $\mu=\cos \theta$. We note that in the previous equation the spatial term is strictly dependent on the electron distance from the center of the star. The Emission-Kinematic term is the integral through the black-body emission weighted by the effective inverse Compton cross section, that also depends on the distance of the star and on the angle $\varphi$, since $z$ (and $x$) depends of $r$ and $\varphi$. In addition, we note that for $r=R$, i.e. $\rho=1$, the spatial term is null by definition~\footnote{The term $(1 + \rho^2 - 2 \rho \mu)^{-3/2}$ is always valid for any $\mu$ in the region of interest $\rho < \mu \leq 1$, except that for $\mu=1$ and $\rho=1$, but since the integral is null by definition in this case we will assume that $\rho<1$  and the effective cross section is null for $\rho=1$.}, and for large distances from the star it reduces to the inverse square law of $r$, as expected.

Assuming that the low energy photons being emitted by a star have a spherical symmetry, we can choose a reference system such that the electron position is along the Z axis (see bottom panel of Fig. ~\ref{fig:SunIC3D}), i.e. $P \rightarrow P'= (0,0,r)$, ($P'= R' ~ P$, where $R'$ is the appropriate rotation matrix). In this system, the electron direction is defined by new polar and azimuthal angles $\hat{p}'=(\varphi'_e, \theta'_e)$, such that $\hat{p}'=R' ~ \hat{p}$.

In addition, again because of azimuthal symmetry, we can also choose a reference system such that the electron direction lies in the X-Z plane, i.e. $\hat{p}'\rightarrow \hat{p}^*=(c^*_x, 0, c^*_z)$. This transformation corresponds to a rotation of an angle $- \varphi'_e$ around the $Z'$ axis, i.e. to the rotation matrix $R_{Z'}(-\varphi'_e)$. In this way
$\hat{p}^* = ( \sin{\theta'_e}, 0, \cos{\theta'_e} )$.

In this system the photon direction is seen by an electron in the position $P'(=P^*)$ with the three direction cosines given by (we still use the notations $\varphi$ and $\theta$ for the azimuthal and polar angles for $Q$):

\begin{equation}
    \hat{\epsilon} = \left( -\frac{R \cos \varphi \ sin \theta}{d}, -\frac{R \sin \varphi \ sin \theta}{d}, \frac{r-R \cos \theta}{d} \right),
    \label{eq:epsdir}
\end{equation}
and the cosine of the angle between the electron and the photon $z$, is:
\begin{eqnarray}
    && z =  \hat{p}^* \cdot \hat{\epsilon} =  \\ \nonumber
    && -\frac{R \cos{\varphi} \sin{\theta}}{d} \sin{\theta'_e} + \frac{(r-R\cos \theta)}{d} \cos{\theta'_e} = \\ \nonumber
    && \frac{- \rho \cos{\varphi}\sin{\theta} \sin{\theta'_e} +(1 - \rho \mu) \cos{\theta'_e}}{\sqrt{1 + \rho^2 - 2 \rho \mu}}. \nonumber
\end{eqnarray}

Therefore the effective cross section can be calculated as a function of the only parameter $\theta'_e$ (polar angle) for the electron direction and the above effective cross section becomes (to keep the notation light we are again dropping the prime for the electron polar angle):

\begin{eqnarray}
    \label{eq:eq43}
     && \sigma(E_e, r, \theta_e) = \\ \nonumber
    && \frac{\rho^2}{4 \pi} \int d\varphi \int_{\rho}^{1} d\mu \frac{\mu - \rho}{(1 + \rho^2 - 2 \rho \mu)^{3/2}} \times \\ \nonumber
    && \int d\epsilon~ n_{ph}(\epsilon)~ (1-\beta z)~ \sigma_C(x).
\end{eqnarray}
For large values of $r$, i.e. $r>>R$, it reduces to:


\begin{equation}
\sigma(E_e, r, \theta_e) \approx \frac{1}{4} \frac{R^2}{r^2}  \int d\epsilon~ n_{ph}(\epsilon)~ (1-\beta z)~ \sigma_C(x),
    \label{eq:eq44}
\end{equation}
with $z = \cos{\theta_e}$. This is expected, since at large distances the star is seen as a half emitting surface with the intensity that decreases as $1/r^2$.

In the case of the $z = \cos{\theta_e}$ the spatial term can be separated by the emission-kinematic one, and can be easily calculated (see also~\cite{Orlando:2008uk}) as:

\begin{eqnarray}
    && \frac{\rho^2}{4 \pi} \int d \varphi\int_{\rho}^{1} d\mu \frac{\mu - \rho}{(1 + \rho^2 - 2 \rho \mu)^{3/2}} = \\ \nonumber
    && \frac{1}{2} \left( 1-\sqrt{1-\rho^2} \right)
\end{eqnarray}
In this case the effective cross section is given by:

\begin{eqnarray}
    \label{eq:eq46}
    && \sigma(E_e, r, \theta_e) \approx \\ \nonumber
    && \frac{1}{2}\left( 1-\sqrt{1-\rho^2} \right) \int d\epsilon~ n_{ph}(\epsilon)~ (1-\beta z)~ \sigma_C(x).
\end{eqnarray}

In the Thomson limit and in the case of head-on collision the Eq.~\ref{eq:eq46} becomes :
\begin{eqnarray}
    \label{eq:eq48}
    && \sigma(E_e, r, \theta_e=\pi) \approx \\ \nonumber 
    && \frac{1}{2} \sigma_T \left( 1-\sqrt{1-\rho^2} \right) (1+\beta) \int d\epsilon~ n_{ph}
    (\epsilon).
\end{eqnarray}

For an isotropic electron distribution the total effective cross section is given by~\footnote{The occultation of the star on the electron trajectories should be also taken into account.}:

\begin{equation}
    \sigma(E_e, r) = \int  
    \sigma(E_e, r, \theta_e) ~ \sin{\theta_e} d\theta_e.
    \label{eq:eq47}
\end{equation}

\begin{figure}[t]
    \centering
    \includegraphics[width=0.49\textwidth, height=0.25\textheight]{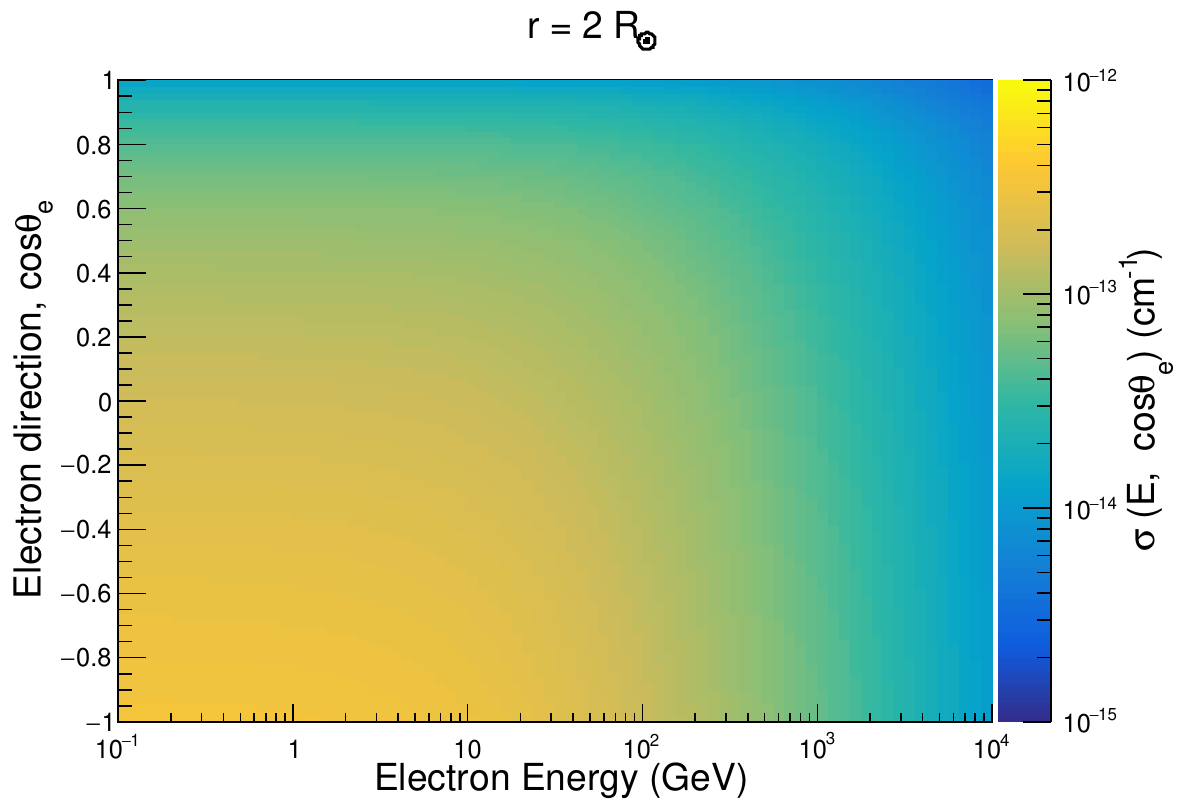}
    \includegraphics[width=0.49\textwidth, height=0.25\textheight]{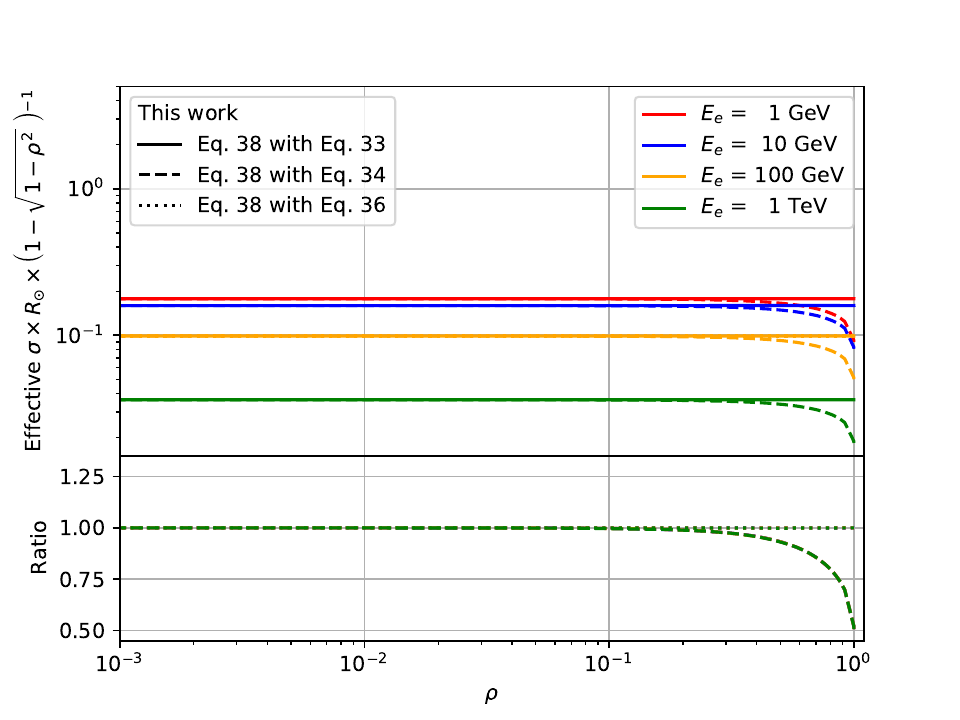}
    \caption{Effective cross section calculated for the solar photospheric temperature of 5770 K. Top panel: Cross section as a function of electron energies and directions at $r=2 R_\odot$ (Eq.~\ref{eq:eq43}). Bottom panel: Total effective cross section multiplied by the sun radius value and divided by $\left( 1-\sqrt{1-\rho^2} \right)$ for an isotropic electron distribution as a function of the inverse of the heliospheric distances $\rho=R_{\odot}/r$ (the Sun is a $\rho=1$). Eq.~\ref{eq:eq43} has been used with the full calculation of Eq.~\ref{eq:eq47} (solid lines) and for the large distances approximation with Eq.~\ref{eq:eq44} (dashed lines). The result obtained with the Eq.~\ref{eq:eq46} is also shown with dotted lines.}
    \label{fig:SunEq43}
\end{figure}

Fig.~\ref{fig:SunEq43} shows the effective cross section for an isotropic electron distribution (Eq.~\ref{eq:eq43})~\footnote{The integral of Eq.~\ref{eq:eq43} has been evaluated with the Genz-Mallik adaptive quadrature integration in multi-dimensions using rectangular regions~\cite{GENZ1980295} implemented in ROOT~\cite{Brun:1997pa} version 6.24/06.} calculated for the solar photospheric temperature of 5770 K. The top panel shows the cross section as a function of the electron energy (horizontal axis) and direction (vertical axis) at the distance of $r=2 R_\odot$. The cross section is higher for negative $\cos{\theta_e}$ since those directions correspond to head-on collisions. The bottom panel shows the cross section as a function of the inverse of the heliospheric distance $\rho=R/r$ for different electron energies. The full calculation by using Eq.~\ref{eq:eq43} is shown with solid lines, and the large distances approximation in Eq.~\ref{eq:eq44} is shown with dashed lines. The two equations agree starting from a distance of about 10 Sun radii, while for shorter distances Eq.~\ref{eq:eq44} underestimates the full prediction. The calculation performed by using  Eq.~\ref{eq:eq46} is shown with dotted lines. The approximation of Eq.~\ref{eq:eq46} agrees  with the full calculation within 1\% or better.

Finally, for a given electron energy $E_e$ the total interaction length for the IC scattering as a function of the distance $r$ and for a given electron direction $c_z$, $\lambda(E_e, r, \theta_e)$ in a stellar black-body photon field is given by:

\begin{equation}
   \lambda(E_e, r, \theta_e) = \frac{1}{\sigma(E_e, r, \theta_e)}.
   \label{eq:eq49}
\end{equation}

\begin{figure}[!th]
    \centering    
    \includegraphics[width=0.49\textwidth, height=0.25\textheight]{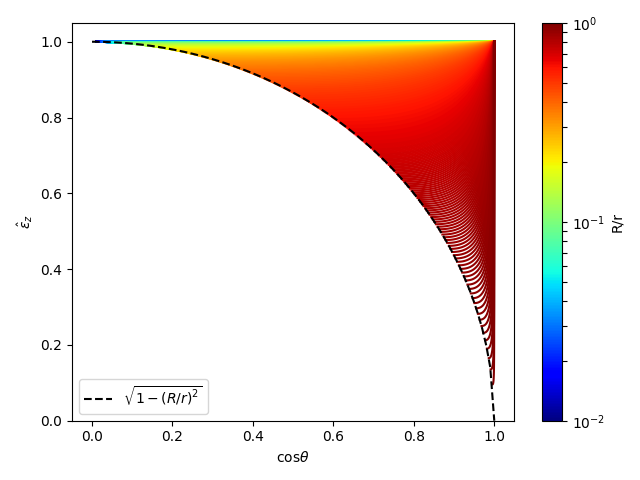}
    \caption{Z-axis component of the emission photon direction unit vector $\hat{\epsilon}_z$ as a function of the cosine of the polar angle for several values of $R/r$ (colored lines). The lower limit as a function of $\cos \theta$ is shown with the black dashed line.}
    \label{fig:epsz}
\end{figure}

The component along the Z axis of the emitted photon direction is $\hat{\epsilon}_z=\frac{r-R\cos \theta}{d}$ and, for a given distance $r$, the cosine of the photon polar angle is in the range $(R/r,1]$. Fig.~\ref{fig:epsz} shows $\hat{\epsilon}_z$ as a function of the cosine of the polar angle for several values of $R/r$. The lower limit as a function of $\cos \theta$ is also shown with the black dashed line, i.e. $(\hat{\epsilon}_z)_{min}=\sqrt{1-R^2/r^2}$. For lower distances, the cosine of the polar angle assumes values close to 1 (i.e. orthogonal to the star in the direction of the electron position), while for larger distances the cosine of the polar angles spans between 0 and 1.



\section{Monte Carlo calculation of the inverse Compton emission from stars}
\label{sec:MCICstar}

With the above calculation in hand, we can now proceed to calculate the IC emission from stars with a Monte Carlo approach.

\begin{figure}[!th]
    \vspace{1cm}
    \centering
    \includegraphics[scale=0.5]{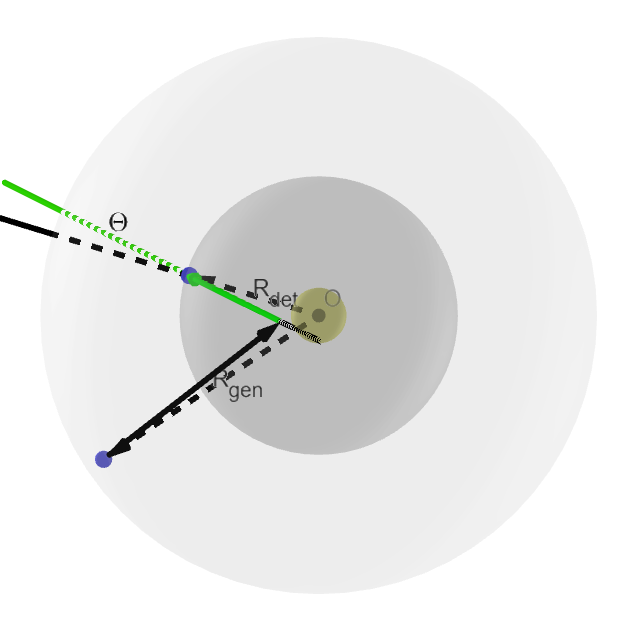}    \caption{Monte Carlo geometry used in the simulation to generate and propagate electrons (black arrows) and to detect IC photons (green arrows); $\Theta$ is the angle between the photon and the normal direction at the detection surface.}
    \label{fig:mcgeo}
\end{figure}

To evaluate the yields of photons produced by the IC scattering of electrons in a low-energy photon field from a star, we have simulated several samples of electrons with different kinetic energies impinging a sphere of radius $R_{gen}$ surrounding the star (see Fig. \ref{fig:mcgeo}), with an isotropic and uniform distribution with a fluence given by $1/\pi R_{gen}^2$ inside the sphere. A secondary photon produced by IC scattering is detected on sphere of radius $R_{det}$ and it is seen with an angle $\Theta$ with respect to the normal direction on that sphere. 

For an electron with energy $E_e$ at the initial position $\vec{r}_0$ with direction $\hat{d_0}$, the path length $s$ from the current position to the site of the next collision is calculated by using the sampling formula $s = min(s_r, s_{max})$, where $s_r= - \lambda(E_e, r, c_z) \log(1-u)$ with $u$ a random number distributed with a uniform PDF in $[0,1[$ and $\lambda(E_e, r, c_z)$ given by Eq.~\ref{eq:eq49}. The electron direction cosine $c_z$ is calculated in the reference system where the electron position is on the Z-axis and its direction lies in the X-Z plane. The following interaction occurs at the position $\vec{r}_{i+1} = \vec{r}_{i} + s \hat{d}_{i}$ where $\hat{d_{i}}$ is direction vector unit. When the sampled value of $s_r$ is less than $s_{max}$, a real interaction is simulated, producing a secondary photon, as discussed in the Sec.~\ref{sec:ICiso}.  
 
The energies of photons emitted from the star are extracted from the black-body radiation spectrum (Eq.~\ref{eq:bbr}), while their directions are extracted according to Eq.~\ref{eq:epsdir} in the reference coordinate system with the electron position on the Z-axis and its direction in the X-Z plane. The azimuthal and polar angles $\varphi$ and $\theta$ are then randomly extracted: $\varphi$ is taken from a uniform distribution from 0 to 2$\pi$, while $cos{\theta}$ is taken from a uniform distribution from $\rho$ to 1. The scattered photon and the scattered electron are then rotated back to the master coordinate reference system. This procedure continues until the electron leaves the detection sphere or hits the star surface.

The yield of secondary photons produced from CREs, $Y(E_{\gamma} | E_k, \Theta)$, is calculated by counting the secondary photons which escape from the detector surface (i.e. the outgoing photons) as:

\begin{equation}
Y(E_{\gamma} |~ E_k, \Theta) = \frac{N(E_{\gamma} |~ E_k, \Theta)}
{N_e(E_k) \Delta E_{\gamma}}
\label{eq:yieldtheta}
\end{equation} 
where $N_{e}(E_k)$ is the number of primary electrons generated with kinetic energy $E_k$ and $N(E_{\gamma}|~E_k, \Theta)$ is the number of secondary photons escaping from the detection surface with energies between $E_{\gamma}$ and $E_{\gamma} + \Delta E_{\gamma}$ and with angles between $\Theta$ and $\Theta + \Delta \Theta$.

The flux of secondary photons seen by an observer on the detection sphere (in units of $\units{GeV^{-1}~cm^{-2}~s^{-1}}$) in an angular range $[\Theta$, $\Theta + \Delta \Theta]$ is given by:

\begin{eqnarray}
\label{eq:flux}
&& \phi(E_{\gamma}, \Theta) = \\ \nonumber && \cfrac{1}{4 \pi R_{det}^{2}}~
\int Y(E_{\gamma} | E_k, \Theta) ~ (4 \pi) (\pi R_{gen}^{2}) I_{e}(E_k)~dE_k = \\ \nonumber
&& \pi \cfrac{R_{gen}^{2}}{R_{det}^{2}}~
\int Y(E_{\gamma} | E_k, \Theta) ~ I_{e}(E_k)~dE_k
\end{eqnarray}
where $(4 \pi) (\pi R_{gen}^{2}) I_{e}(E_k)$ is the total number of electrons entering in the simulated volume in units of $\units{GeV^{-1}~s^{-1}}$. We remark here that photons hitting the star surface are killed and then they are not able to reach the detection surface. 

The differential intensity spectrum of secondary photons with an angle $\Theta$ from the star's center (in units of$\units{GeV^{-1}~cm^{-2}~sr^{-1}~s^{-1}}$) emitted from IC scatterings is given by:
 
\begin{equation}
I (E_{\gamma}, \Theta) = \frac{\phi(E_{\gamma}, \Theta)}{\Delta \Omega(\Theta)}
\label{eq:intensity}
\end{equation}
where $\Delta \Omega(\Theta)$ is the solid angle corresponding to a $\Theta$ angle in the range $[\Theta, \Theta + \Delta \Theta]$, i.e. $\Delta \Omega(\Theta)=2 \pi \Delta \cos{\Theta}$. 

\section{Solar inverse Compton emission}

Here we present the photon emissivity of the Sun by IC scattering. We assume the temperature of the Sun surface of 5770 K, the solar radius of $R_\odot=6.9551 \times 10^{10}\units{cm}$, the electron generation sphere with a radius of 1 AU and the detection sphere again with a radius of 1 AU, i.e. $R_{gen}=R_{det}= 1\units{AU}$. In this way, we assume that the CRE intensity~\footnote{We use the total intensity of electrons and positrons, and we refer to them as electrons.} at the generation sphere is the same observed at the Earth by the Fermi-LAT~\cite{Fermi-LAT:2009yfs,Fermi-LAT:2017bpc}, AMS-02~\cite{Aguilar:2014fea} and DAMPE~\cite{Ambrosi:2017wek}. We have extrapolated the data down to 0.1\units{GeV} and up to 10\units{TeV} as we did in our previous work~\cite{Mazziotta:2020uey}.

We have simulated the interactions of electrons impinging on the generation sphere in a wide energy range from 0.1\units{GeV} to 10\units{TeV}, while the energy of secondary particles has been simulated down to 10\units{eV}. The primary kinetic energy values are taken on a grid of $80$ equally spaced values in a logarithmic scale. 

We have simulated the interactions without and with an interplanetary magnetic field, with solar winds up to the Sun surface, i.e. we are not considering the strong and irregular inner magnetic field near the Sun~\cite{Mazziotta:2020uey}.
  
In the case without magnetic field, the electron trajectories are straight lines, unless an IC scattering occurs, where the scattered electron can change its direction. 

In the magnetic field case, we are assuming that the interplanetary magnetic field (IMF) is described by the Parker model~\cite{Parker:1958zz} for $r>R_{\odot}$. The three components of the IMF are given by~\cite{Mazziotta:2020uey}:

\begin{eqnarray}
B_{r} & = & \pm f B_{E} \left( \frac{R_{E}}{r} \right)^{2} \label{eq:br} \notag \\
B_{\theta} & = & 0 \\
B_{\phi} & = & -B_{r} \tan \xi \notag
\end{eqnarray}
The polar component $B_{\theta}$ is assumed always null. The angle $\xi$ is defined as:

\begin{equation}
\tan \xi(r, \theta) = \frac{\omega_{S} \left( r - R_{\odot} \right) \sin \theta}{v_{SW}}      
\end{equation}
where $\theta$ is the polar angle, $\omega_{S}=2.69 \times 10^{-6} \units{rad/s}$ is the angular velocity of the Sun (corresponding to a period of about $27 \units{days}$) and $v_{SW}$ is the velocity of the solar wind (its typical value is $400\units{km/s}$). At the distance $R_{\odot}$ the components $B_{\phi}$ is also  null, i.e. the B-field i radial at the Sun surface.

In the previous equations the intensity of the field $B_E$ is given by $B_{E}=B_0/\sqrt{1+\tan^2 \xi(R_E,\pi/2)}$, where $B_{0}$ is the intensity of the magnetic field at the Earth (its typical value is about $5\units{nT}$), and $R_{E}=1\units{AU}$ is the Sun-Earth distance. The constant $f$ is given by:

\begin{equation}
f = 1 - 2H(\theta - \theta')    
\end{equation}
where $H$ is the Heaviside function and the angle $\theta'$ is the polar position of the heliospheric current sheet (HCS) defined as:

\begin{equation}
\theta' = \frac{\pi}{2} - \arctan \left[ 
\tan \alpha  \sin \left( \phi + \frac{\omega_{S} \left( r - R_{\odot} \right)}{v_{SW}} \right)
\right]
\end{equation}
where $\phi$ is the azimuth angle and we have indicated with $\alpha$ the tilt angle, i.e. the maximum latitude of the HCS; finally, the $\pm$ sign in Eq.~\ref{eq:br} depends on the polarity of the magnetic field. 

We assume a constant speed of the solar wind with a radial direction, i.e. $\vec{v}_{SW}(\vec{r}) = v_{SW} \, \hat{r}$. The electric field can be written as the gradient of a potential function:

\begin{equation}
\vec{\Ef} (x,y,z) = -\nabla V(x,y,z)~.
\label{eq:gradient}
\end{equation}
Assuming that the potential vanishes at the 
heliospheric equator $z=0$, the electric potential $V$ is given by~\cite{lipari2014solar}:

\begin{equation}
V(x,y,z) = 
\mp f \, B_0 \;
\omega_S \, R_E^2\; \frac{z}{r} ~
\label{eq:vpotential}
\end{equation}
where the $\mp$ sign is valid above (below) the HCS.

The energy variation of a charged particle due to the electric potential, $\Delta E$, is given by:

\begin{equation}
 \Delta E = q \left [ V(\vec{r}_f) - V(\vec{r}_i)\right ]
\label{eq:potential1}
\end{equation}
and is proportional to the difference between the electric potential calculated at the initial and final points $\vec{r}_i$ and $\vec{r}_f$ on the trajectory.

To keep the discussion as simple as possible, in this work we assume a null tilt angle, a positive polarity, $B_0 = 5$ nT and $v_{SW} = 400\units{km/s}$. The magnetic field and the solar wind are considered to be stationary during the simulation. In the case of magnetic field, the tracking of one step is performed combining the helix approximation and an adaptive Runge-Kutta-Nystroem algorithm~\cite{Brun:1987ma} due to the magnetic field configuration. If the travel angle at the beginning of a step is larger than $\pi/2$ the helix approximation is used. Then, the energy variation due to the solar wind of the charged particle is calculated with Eq.~\ref{eq:potential1}, where the final position is the one calculated after a single step.

We have simulated $10^7$ and $5 \times 10^6$ electrons for each incoming energy, in the cases without and with magnetic field, respectively. The total statistics in the configuration with magnetic field is half of the statistics in the configuration without magnetic field, since the former case requires a higher CPU consumption. However, since the yield is highly enhanced near the Sun, the statistics on the IC gamma rays still quite high.

\begin{figure}[!t]
    \centering
    \includegraphics[width=0.49\textwidth, height=0.25\textheight]{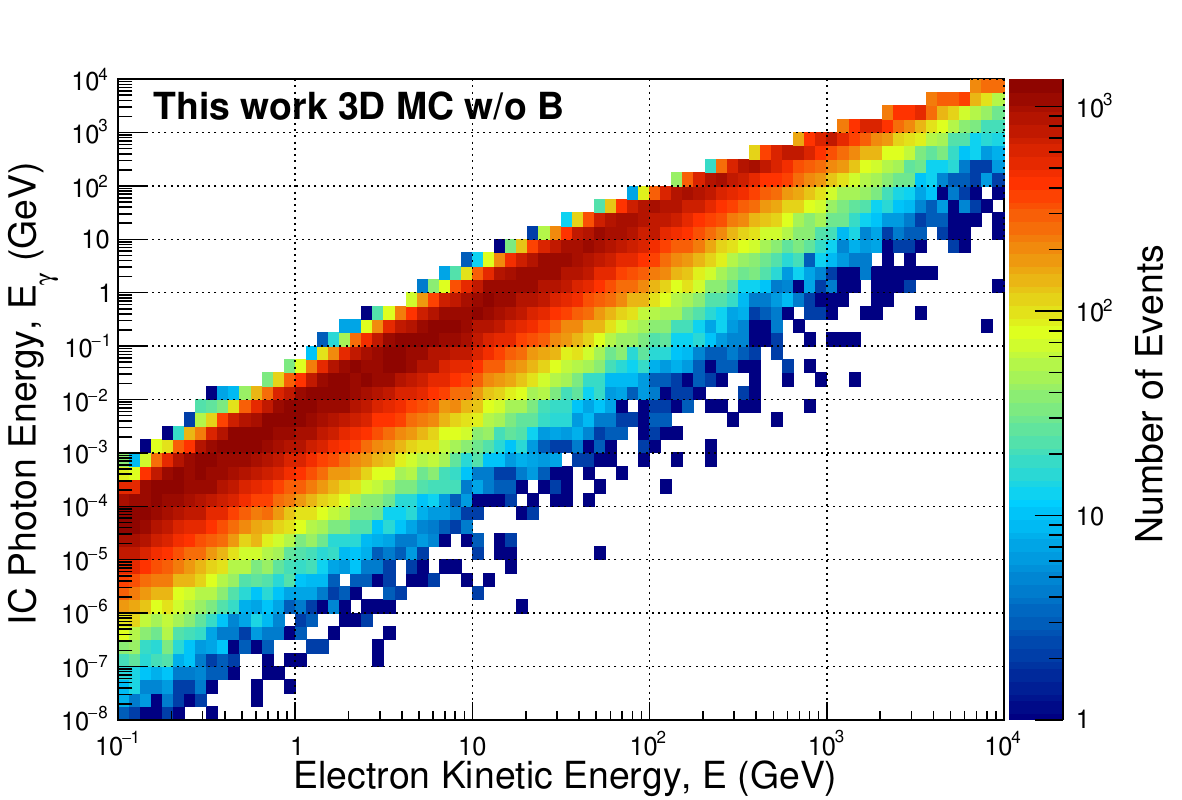}\caption{Distribution of the IC photon energies as a function of the incoming electron energies (configuration without magnetic and electric fields). The plot is obtained after integration on the angular separation of the photons from the Sun.}
    \label{fig:icEphvsEe}
\end{figure}

Fig.~\ref{fig:icEphvsEe} shows the IC gamma-ray energies as a function of the incoming electron energies at the generation surface in the configuration without magnetic field. The plot is obtained after integration on the angular separation of the photons from the Sun. The photons with energy above 10 MeV are mainly produced by electrons above 1 GeV.

\begin{figure}[!t]
    \centering
    \includegraphics[width=0.49\textwidth, height=0.25\textheight]{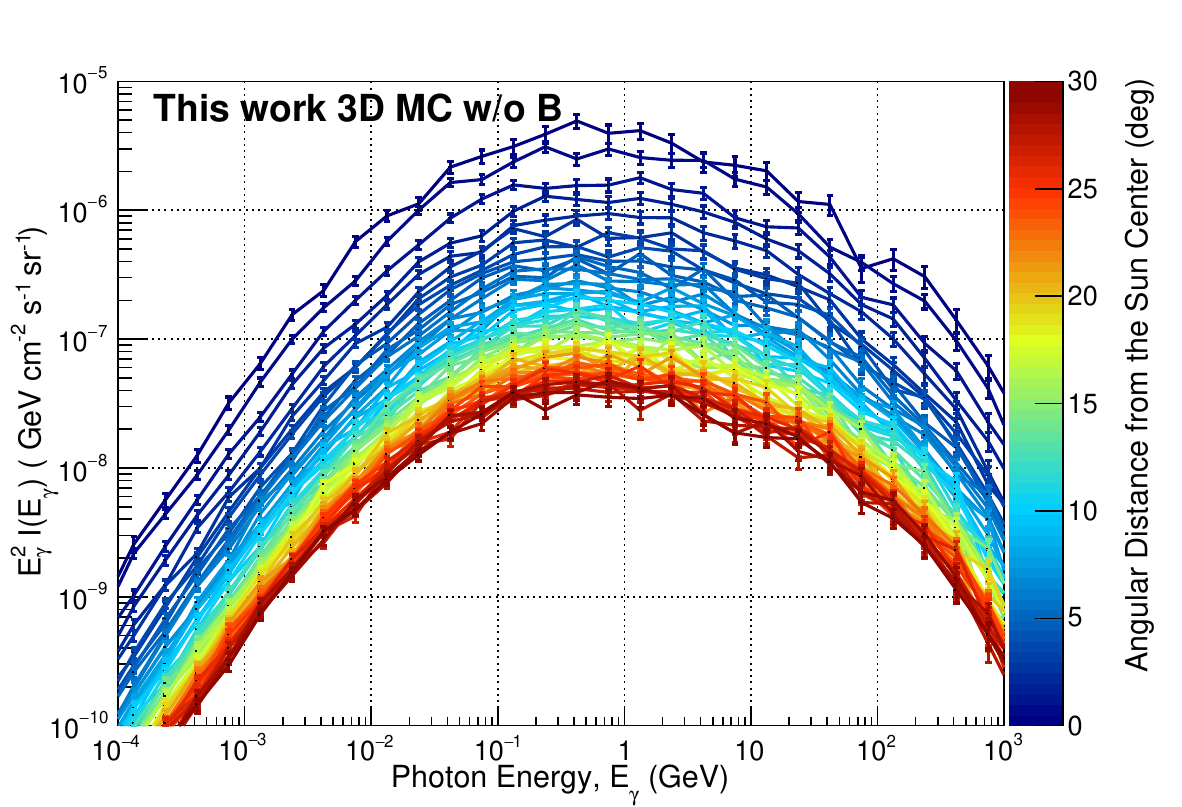}
    \caption{Calculated IC spectral intensity for various angular distances from the Sun up to $30\degrees$ with step of $0.5\degrees$. We show the spectral intensity  multiplied for $E^2_{\gamma}$ as a function of photon energy. The darker blue line corresponds to the interval $[0.0\degrees, 0.5\degrees]$, while the darker red line shows the intensity in the interval $[29.5\degrees, 30.0\degrees]$. The error bars represent the statistic error due to the number of simulated events. }
    \label{fig:icE2IntnoB}
\end{figure}

Fig.~ \ref{fig:icE2IntnoB} shows the calculated IC spectral intensity for various angular distances from the Sun up to $30\degrees$, with a step of $0.5\degrees$, again in the case without magnetic field. We show the spectral intensity (multiplied by $E^2_{\gamma}$) as a function of photon energy.

\begin{figure*}
\centering
\includegraphics[width=0.49\textwidth, height=0.25\textheight]{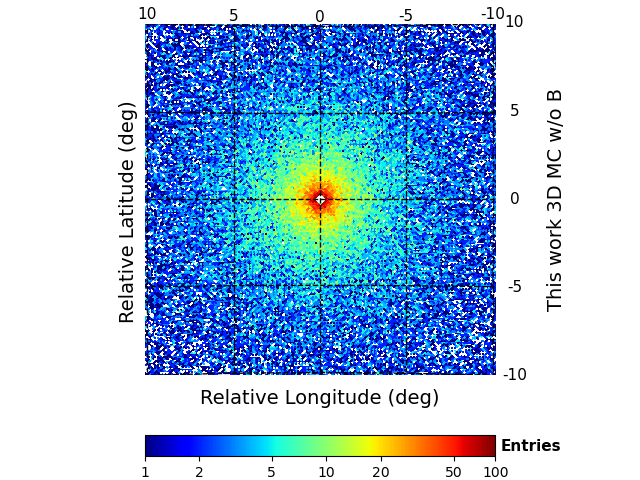}
\includegraphics[width=0.49\textwidth, height=0.25\textheight]{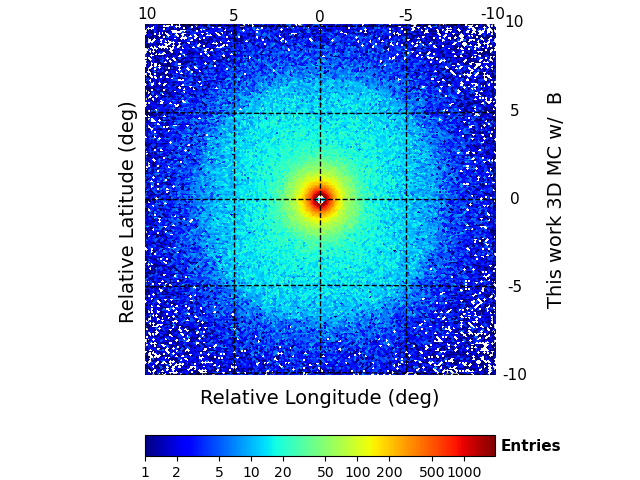}
\includegraphics[width=0.49\textwidth, height=0.25\textheight]{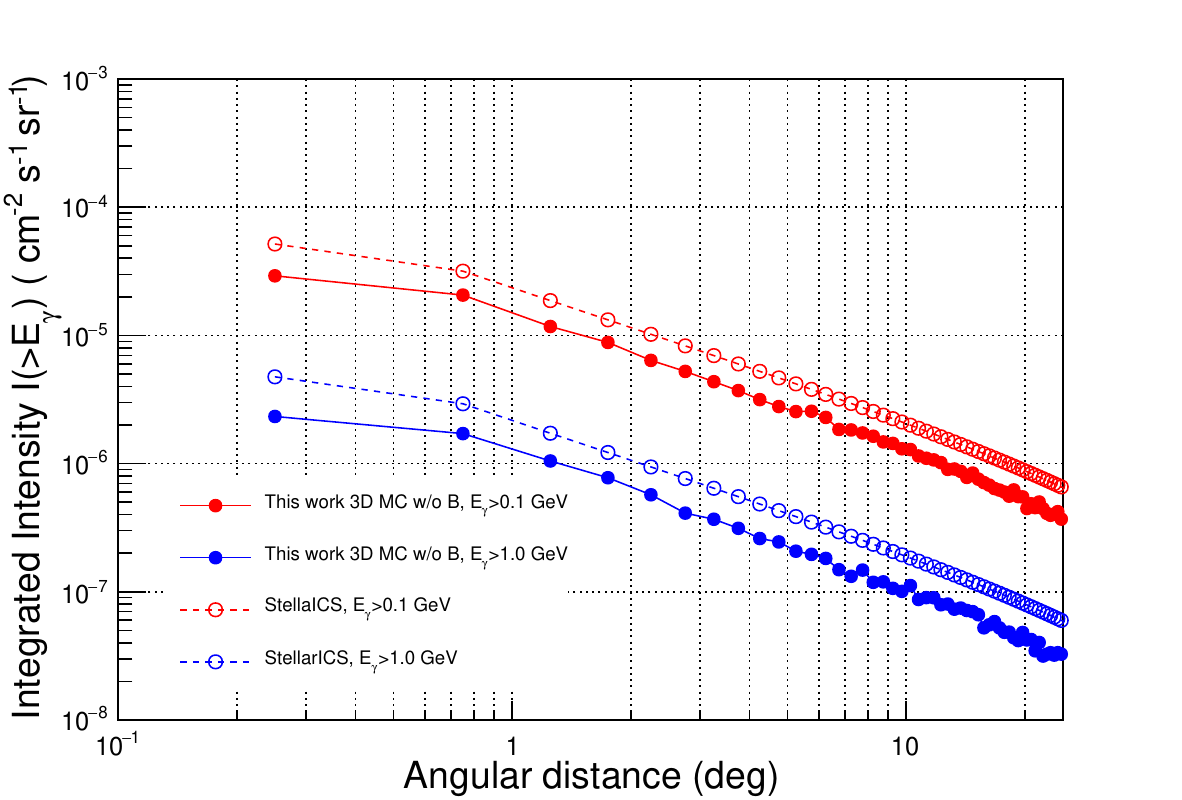}
\includegraphics[width=0.49\textwidth, height=0.25\textheight]{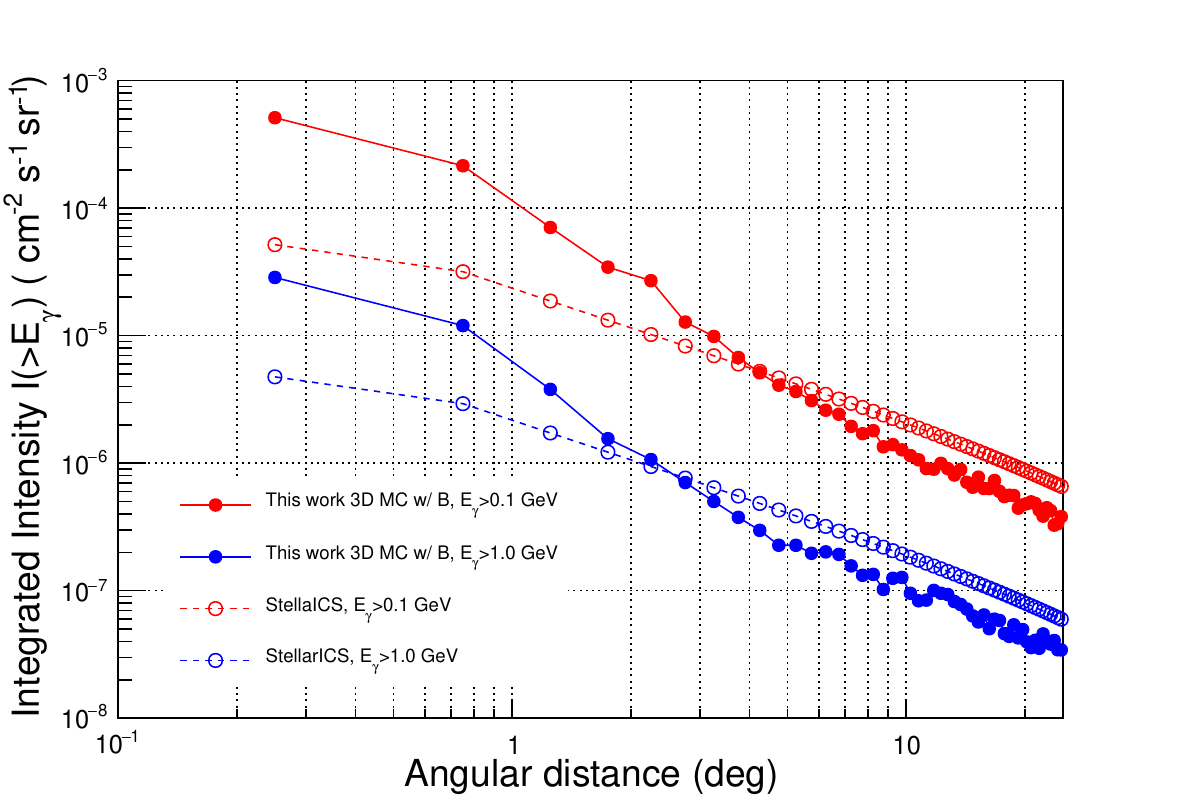}
\includegraphics[width=0.48\textwidth, height=0.25\textheight]{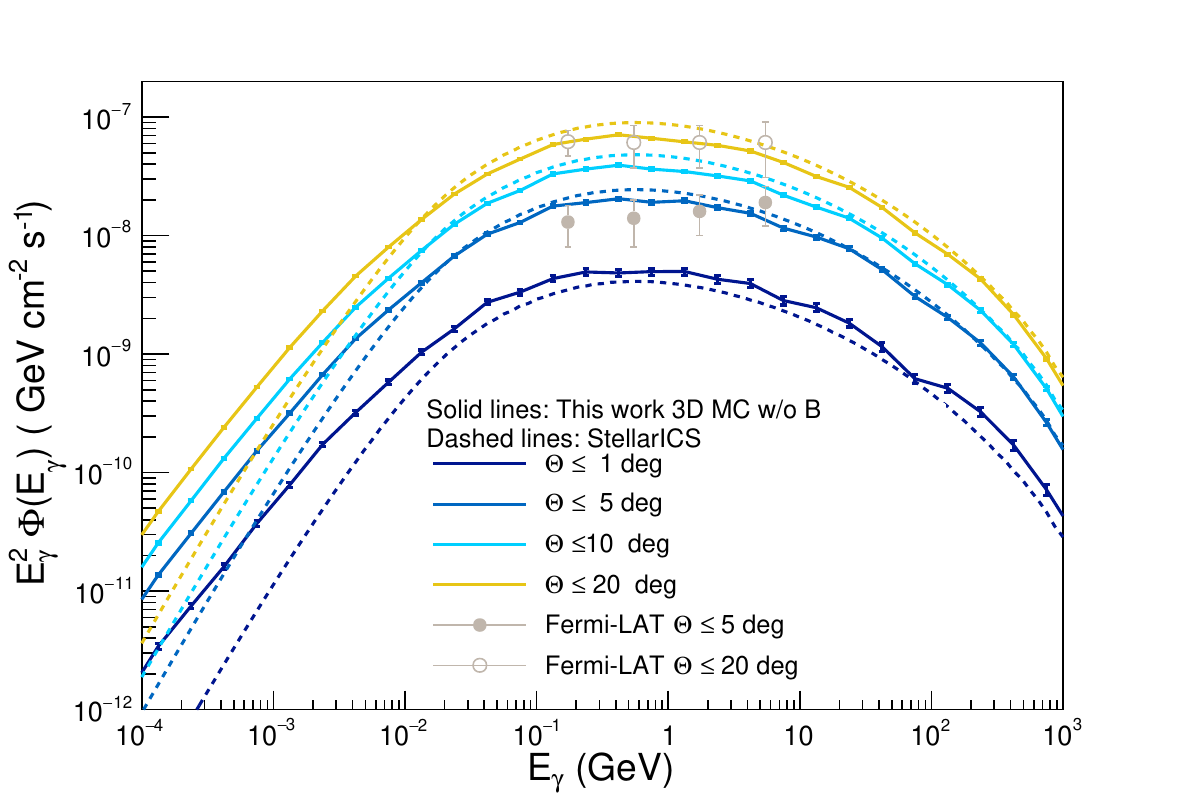}
\includegraphics[width=0.48\textwidth, height=0.25\textheight]{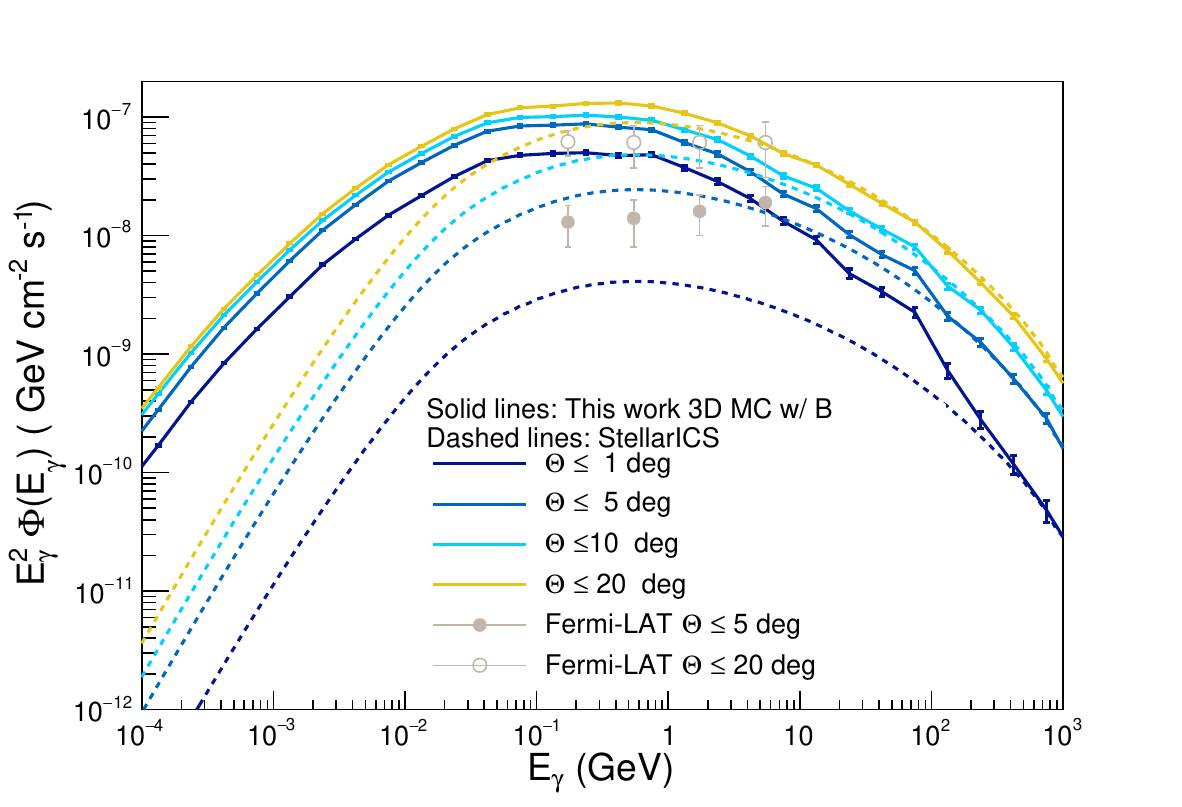}
\caption{Comparison without (left column) and with magnetic and electric field configurations (right column).
Top row: Energy integrated IC photon spatial map around the Sun. The map is built with the HEALPix pixelization scheme with $N_{side}=512$ (pixel area of about 3.995 $\times 10^{-6}\units{rad}$) . Middle row: Intensity profile integrated above 100 MeV (red points) and above 1 GeV (blue points) as a function of the angular distance from the Sun. Solid lines (and filled points) represent the results of this work, while dashed lines (and open points) represent the StellarICS~\cite{Orlando:2020ezh} results calculated with same configuration. Bottom row: Calculated IC spectral flux (multiplied for $E^2_{\gamma}$) without magnetic field as a function of the photon energy, integrated for angular distances from the Sun center within (from bottom to top) $1\degrees$, $5\degrees$, $10\degrees$ and $20\degrees$ respectively. Dashed lines represent the StellaICS~\cite{Orlando:2020ezh} results calculated with same configuration. Grey points show the Fermi-LAT measurements~\cite{Abdo:2011xn}.}
\label{fig:noBandB}
\end{figure*}

Fig.~\ref{fig:noBandB} shows the results of this simulation in the configuration without (left column) and with magnetic field (right column). In the top row, we show the energy integrated gamma-ray spatial map around the Sun produced by IC interactions built with the HEALPix pixelization~\cite{Gorski_2005}~\footnote{HEALPix website – currently \url{http://healpix.sourceforge.net} or \url{https://healpix.sourceforge.io.}}. The emission is mainly located near the Sun and extends far away from it for a tens degrees. In the case of magnetic field the emission is much narrow towards the Sun.

In the middle row of Fig.~\ref{fig:noBandB}, we show the intensity profiles integrated above 100 MeV (red lines and points) and above 1 GeV (blue lines and points) as a function of the angular distance from the Sun. The results of this work are compared with those obtained with the StellarICS code~\cite{Orlando:2020ezh} with the same configuration. The calculation of the IC intensity with the StellarICS code is performed integrating the IC emission along the line-of-sight direction in a plane with respect to the observer at a given angular distance with respect the center of the star (see also~\cite{Lai:2022qif}). In the case without magnetic field, the predicted flux from our 3D Monte Carlo procedure is slightly lower than the predicted flux from StellarICS in all angular windows. The shape is very similar and in the integrated intensity decreases with the angular distance, and can be represented with a power-law $I(\Theta) \propto \Theta^{-1}$. In the case of magnetic field, our results are above the StellarICS ones for $\Theta<4\units{deg}$ and the integrated intensity is much steeper.

Finally, in the bottom row of Fig.~ \ref{fig:noBandB}, we show the calculated IC spectral flux (multiplied for $E^2_{\gamma}$) as a function of the photon energy, integrated for angular distances from the Sun center within (from bottom to top) $1\degrees$, $5\degrees$, $10\degrees$ and $20\degrees$ respectively. Dashed lines represent the StellarICS~\cite{Orlando:2020ezh} results calculated with the same configuration. Grey points show the Fermi-LAT measurements~\cite{Abdo:2011xn}. Our results agrees with the StellarICS predictions without magnetic field, while in the case of magnetic field the flux is higher in the region close to the Sun. We also see that the predicted flux from our simulation is much higher at low gamma-ray energies ($< 100\units{MeV})$. 

\section{Conclusions and outlook}
\label{sec:conc}

In this work we have considered the IC scattering between energetic electrons and low energy photon fields, and we have derived the differential cross section in a contest of 3D Monte Carlo approach. This process is frequently used in high-energy astrophysics for the evaluation of IC photon emission, in particular for the diffuse solar emission. Our study also represents a step forward in understanding the solar gamma-ray emission. 

In our formalism we are not implemented any approximation, although sometimes we have also shown the results in some limits as for example in the Thomson regime. The approach shown in this work can be also used in presence of magnetic and electric fields.

We have calculated the solar IC emission due to galactic CREs without and with interplanetary electromagnetic field. In the case without magnetic field, we find that our results agree with those from the StellarICS calculation~\cite{Orlando:2020ezh}. However, we see some deviations at photon energies below 100 MeV, as already noted in ref.~\cite{Lai:2022qif}, although at much lower energy. 

We have also implemented a configuration with an interplanetary magnetic field described by the Parker model to calculate the solar diffuse component. In this case, although we consider a simple scenario, we see an increase of the emission close to the Sun direction, probably due to the an increase of the time spent by electrons in the radiation field in presence of the magnetic field. If this effect will be confirmed considering a more accurate solar magnetic configuration that also includes the magnetic field near the Sun, it will allow to establish a better modeling of the solar diffuse emission.

The Sun is moving in the galactic reference frame, spanning a latitude interval of about $\pm 60\degrees$. The maximum (minimum) latitudes are reached during spring (fall) and the corresponding longitudes are of about $\pm (60 - 70)$\units{degrees}. On the other hand, during winter (summer), the Sun reaches the minimum distance of a few degrees from the galactic center (anti-center). The isotropic diffuse gamma ray background measured by the Fermi-LAT above 100 MeV is about \mbox{$7.2 \times 10^{-6}\units{cm^{-2}~s^{-1}~sr^{-1}}$}~\cite{Fermi-LAT:2014ryh}, that is less than the solar IC up to a few degrees (see Fig.~\ref{fig:noBandB} middle panels). For the galactic interstellar emission, the intensities above 100 MeV are of about \mbox{$4.8 \times 10^{-6} \units{cm^{-2}~s^{-1}~sr^{-1}}$} at high latitudes, \mbox{$9.1 \times 10^{-4} \units{cm^{-2}~s^{-1}~sr^{-1}}$} towards the galactic center and \mbox{$6.0 \times 10^{-5}\units{cm^{-2}~s^{-1}~sr^{-1}}$} towards the galactic the galactic anti-center center respectively~\cite{Fermi-LAT:2016zaq}. Consequently, the solar IC halo should be easily resolved at high galactic latitudes.

It is worth highlighting that high energy gamma-rays produced in the IC scattering can be absorbed through the scattering photon-photon to electron-positron interaction, where the target is provided by the same solar radiation field as discussed in the paper. By assuming the average value of solar energy photons of about $\epsilon$=1.2 eV the IC gamma-ray energy threshold to produce an $e^+ e^-$ pair is about 217 GeV ($E_{th} \approx m^2_e/\epsilon$). So, this process affects the high energy IC photons since they could be absorbed from the production point to the observer. 
In fact, above threshold the cross section increases very rapidly reaching a maximum of $\sigma_{max} \approx 0.26 ~ \sigma_T$ at the value of $x=\frac{\epsilon E_\gamma}{2 m_e^2}(1-\cos{\theta_{\gamma \gamma}}) \approx 2$, then the cross section decreases monotonically, and for large value of $x$ it takes the asymptotic behavior $\sim log(x)/x)$~\cite{Vernetto:2016alq}.
The full calculation could just follow the same formalism introduced in the section V (from Eq.~\ref{eq:eq30}) where we should put the gamma-gamma pair cross section instead of the Compton one, i.e.

\begin{equation}
   \sigma(E_\gamma, r, \hat{k}) = \int dn ~ (1-\cos{\theta_{\gamma \gamma}})~\sigma_{\gamma \gamma}(x).
   \label{eq:eq30_1}
\end{equation}
The attenuation of gamma ray can be calculated integrating along the line of flight of the photons up the observer and this could be done at posterior since we can assume that the photons move along a straight line (see also Ref. ~\cite{Vernetto:2016alq}).

The IC model discussed in this work is now implemented in the latest version of the {\tt FLUKA} simulation code~\cite{Ferrari:2005zk, FLUKA1, Arico:2019pcz,Ballarini:2024uxz} and allows to better calculate the gamma-ray emission in complex geometries, even in presence of magnetic and electric fields, also including synchrotron radiation emission along particle trajectories. The results achieved with {\tt FLUKA} will be published in a separate work.

\begin{acknowledgments}
We thank Francesco Loparco and Alfredo Ferrari for helpful comments and discussions during the preparation of the manuscript.

Some of the results in this paper have been derived using the HEALPix (K.M. Górski et al., 2005, ApJ, 622, p759) package.

In this work we used custom software based on Fortran, C++, Python~\cite{van1995python,Hunter:2007} and ROOT toolkit~\cite{Brun:1997pa}. GeoGebra is also used for some images.
\end{acknowledgments}

\bibliographystyle{apsrev4-2}
\bibliography{MC_IC.bib}{}

\end{document}